\documentclass[reprint]{revtex4-2}

\usepackage{graphicx}% Include figure files
\usepackage{hyperref}
\usepackage{dcolumn}% Align table columns on decimal point
\usepackage{bm}% bold math
\usepackage{braket}
\usepackage{amsmath}
\usepackage{amssymb}
\usepackage{subcaption}
\usepackage{algorithm}
\usepackage{algpseudocode}
\usepackage{makecell}
\usepackage[dvipsnames]{xcolor}

\usepackage{ulem}

\begin{document}

\preprint{APS/123-QED}

\title{Gradient-descent methods for scalable quantum detector tomography}

% using phantom for anonymization 
%\phantom{
\author{Amanuel Anteneh}
\email[Corresponding author: ]{asa2rc@virginia.edu}

\affiliation{Department of Physics, University of Virginia, 382 McCormick Rd, Charlottesville, VA 22903, USA}
%}
%\phantom{
\author{Olivier Pfister}
\email{olivier.pfister@gmail.com}
\affiliation{Department of Physics, University of Virginia, 382 McCormick Rd, Charlottesville, VA 22903, USA}
\affiliation{Charles L. Brown Department of Electrical and Computer Engineering, University of Virginia, 351 McCormick Road, Charlottesville, VA 22903, USA}
%}
\date{\today}

\begin{abstract}
We present a technique for performing quantum detector tomography (QDT) of phase insensitive quantum detectors, a category under which many detectors of interest fall under, using gradient descent-based optimization to learn the positive operator-valued measure (POVM) that best describes the data collected using the detector under study. We numerically benchmark our method against constrained convex optimization (CCO) and show that it reaches higher or comparable reconstruction fidelity in much less time even in the presence of noise and limited probe state resources. We also present a possible extension of our approach to the phase sensitive case via a parametrization of POVMs on the complex Stiefel manifold which enables gradient based optimization restricted to this manifold.
\end{abstract}

\maketitle

\section{Introduction}

Fast and accurate characterization of quantum states via quantum state tomography (QST)~\cite{Smithey1993,Banaszek1996,Wallentowitz1996,Paris2004} and quantum processes via quantum process tomography (QPT)~\cite{Lobino2008, nielsen2021gate} are fundamental tools of quantum information science.  
However, the assumption of well characterized detectors is critical to performing both QST and QPT \cite{feito2009measuring}. Having well characterizing detectors is also crucial for other quantum mechanical experiments ranging from testing predictions of quantum mechanics by observing violations of Bell's inequality \cite{kuzmich2000violation} to measurement-based quantum computing \cite{Raussendorf2001,briegel2009measurement}. Therefore the development of techniques to accurately characterize quantum detectors, called quantum detector tomography (QDT) or quantum measurement tomography, is a key priority. 

Current QDT protocols cast the reconstruction as a constrained convex optimization (CCO) problem. Physicality conditions, such as completeness and positive semidefiniteness, are imposed through constraints, and the resulting problem is typically solved using semidefinite-programming methods \cite{lundeen2009tomography, zhang2012recursive, natarajan2013quantum, schapeler2020quantum, schapeler2021quantum, liu2023optimized, cattaneo2023self, 
schapeler2024scalable}. 
These methods work well for moderate system sizes and in the low data regime. However, they become computationally inefficient for large system sizes due to the high time and space complexity of many semi-definite programming algorithms. 
There has also been work which uses constrained least squares (CLS) methods to perform QDT which offer efficiency improvements~\cite{wang2021two, xiao2022optimal, xiao2023regularization}. 
%These methods require the inversion of a matrix which can be a costly operation for large systems and requires the number of probe states to scale quadratically with the Hilbert space dimension for the inverse to exist~\cite{wang2021two}.}
% This method involves computing matrix inverses which can be expensive as well as an additional projeciton step to ensure the POVM elements are positve semidefinte. Also it seems that the number of probe states D must be at least as great as M^2 (wang2021two says this must be the case for the matrix inverse the method computes to exist) where M is the Hilbert space dimension. This is in contrast to our apporach. For most of our experiemnts we set M=200 but we use D=2000 probes states which is much less than 200^2.

Gradient based methods are the workhorse of much of modern machine learning and statistical learning, such as the backpropagation algorithm used for the training of deep neural networks \cite{lecun2015deep} and the Hamiltonian Monte Carlo algorithm for Bayesian inference \cite{hoffman2014no}, in particular with the advent of automatic differentiation which allows for accurate and efficient computation of gradients \cite{baydin2018automatic, goodfellow2016deep, bishop2023deep}. 
Recently, unconstrained gradient based optimization algorithms, such as gradient descent, have been utilized to solve problems within quantum information such as state preparation and gate synthesis \cite{arrazola2019machine, miatto2020fast, kudra2022robust, yao2024riemannian} as well as state and process tomography \cite{bolduc2017projected, ahmed2023gradient, wang2024efficient, hsu2024quantum, gaikwad2025gradient}. 
In contrast to CCO these methods are typically much simpler and efficient in terms of time and memory complexity. Here we apply such a method to the reconstruction of the positive operator-valued measure (POVM) of phase insensitive quantum detectors. We also study the robustness of our method to errors in the preprocessing step of the tomography process, i.e. acquiring the relevant measurement statistics, and limited data availability. 
Finally we present a possible extension of our approach to the general case of a phase sensitive detector which enables a rank-controlled ansatz for the POVM elements being reconstructed.

\section{Quantum Detector Tomography}
In this section we recall the standard notation and definition of QDT and present our approach for performing QDT using the gradient descent algorithm. Note that the methods developed here are applicable, in principle, to any quantum detector for qubit or qumode based systems but we focus on the latter for this work.

\subsection{General quantum detector tomography}
In quantum mechanics the most general measurement can be mathematically modeled with a POVM \cite{Helstrom1976,benenti2019principles}. A POVM $\{E_n\}_{n=1}^N$ consists of a set of $N$ Hermitian operators $E_n \in \mathbb{C}^{M\times M}$, where $M$ is the dimension of the Hilbert space, that satisfy the following positive semi-definiteness and completeness conditions
\begin{align} \label{povm_constraints}
    E_n \succeq 0 \quad \forall n \\
    \sum_{n=1}^{N}E_n = \mathbb{I}_M 
\end{align}
where $ \mathbb{I}_M$ is the $M \times M$ identity.
Conventional QDT seeks to find a POVM that describes a quantum measurement with $N$ outcomes from experimental data $P$ generated using a set of $D$ tomographically complete  probe states $\{\rho_1, \dots, \rho_D\}$. The common choice of probe states, for continuous variable systems, is the set of coherent states which are tomographically over-complete \cite{zhang2012recursive}. These states are easily prepared experimentally, describing the light emitted by a stabilized laser well over threshold, and can be written in the Fock basis as

\begin{equation}
    \ket{\alpha} = \sum_{k=0}^\infty e^{-\frac{|\alpha|^2}{2}}\frac{\alpha^k}{\sqrt{k!}}\ket{k}
\end{equation}
which leads to Poissonian photon number statistics that are preserved under photon number loss (of course, the average and standard deviation do scale with loss). For these reasons these states serve as excellent probe state resources.
The CCO problem can now be written as

\begin{equation}\label{full_op_prob}
\begin{aligned}
    \min_{E_1,\cdots,E_N} \quad & \sum_{i=1}^{D}\sum_{j=1}^{N}\big(P_{ij} - \textrm{Tr}\big[E_j\rho_i\big]\big)^2 \\
\textrm{subject to} \quad & E_j \succeq 0 \quad \forall j \\
  &\sum_{j=1}^{N}E_j = \mathbb{I}_M 
\end{aligned}
\end{equation}
where $P_{ij}$ is the experimentally measured probability of the measurement outcome associated with $E_j$ computed after measuring many copies of probe state $\rho_i$. Note that since each $E_j$ is an $M \times M$ matrix there are $NM^2$ free variables to optimize in Eq. \ref{full_op_prob}.

\subsection{Tomography of phase insensitive detectors}
If one knows the detector under consideration is phase insensitive then the POVM elements $E_n$ can be expressed as diagonal matrices in some orthonormal basis. 
For example, for detectors which measure photon number the POVM elements can be written in the Fock basis
\begin{equation}
    E_j = \sum_{i=0}^{M-1} \Pi_{ij} \ket{i}\bra{i}
\end{equation}
where $\Pi$ is a real $M \times N$ matrix whose $j$-th column holds the diagonal elements of POVM element $E_j$.
The optimization problem in Eq. \ref{full_op_prob} can then be reduced to one which includes estimation of only $MN$ variables~\cite{lundeen2009tomography},

\begin{equation}\label{diag_op_prob}
\begin{aligned}
    \min_{\Pi} \quad  & || P - F\Pi ||_F^2\\
\textrm{subject to} \quad & \Pi_{ij} \geq 0 \\
  &\sum_{j=1}^{N}\Pi_{ij} = 1 \quad \forall i \in \{1, \cdots, M\}
\end{aligned}
\end{equation}
where $||A||_F=\sqrt{\sum_{ij}A_{ij}^2}$ is the Frobenius norm and $F \in \mathbb{R}^{D \times M}$ is a matrix containing Fock state probabilities for each probe with explicit entries

\begin{equation}\label{Poisson_eq}
    F_{ij} = \frac{|\alpha_i|^{2j}}{j!}e^{-|\alpha_i|^2}
\end{equation}
which follow a Poisson distribution due to the fact that the probe states are coherent states.
The new constraints follow from the fact that {\it(i)}, a diagonal matrix is positive semidefinite if and only if its diagonal elements are nonnegative and {\it(ii)}, for a set of diagonal matrices to satisfy $\sum_i E_i = \mathbb{I}_M$ the sum of corresponding diagonal entries must equal 1.
One can solve for the matrix $\Pi$ in Eq. \ref{diag_op_prob} by using linear inversion to compute $\Pi$ as $\Pi=F^{-1}P$. However this will not guarantee the physicality of the resulting solution matrix~\cite{barbera2025boosting} which motivates the use of constrained optimization techniques.

\subsection{Gradient descent based quantum detector tomography}

Gradient descent is an unconstrained optimization algorithm whose high degree of scalability has made it the algorithm of choice for training modern deep learning models which have anywhere from millions \cite{devlin2019bert, sanh2019distilbert} to billions of free variables \cite{brown2020language, raffel2020exploring, touvron2023llama}.
However, its unconstrained nature initially presents as less favorable compared to CCO methods, as the latter facilitates the straightforward enforcement of physicality constraints in optimization problems. Soft constraints can be incorporated into gradient descent by adding a term to the objective function that penalizes deviations from some physicality condition e.g. adding a penalty term proportional to $(1-\textrm{Tr}[\rho])^2$ when reconstructing a density matrix for QST.
However, these penalty terms do not provide a strict guarantee of physicality of the resulting solution which would be highly desirable.

In our case the constraints in Eq. \ref{diag_op_prob} can be interpreted as requiring the rows of $\Pi$ to be probability vectors.
This constraint can easily be enforced in our gradient descent approach by use of the softmax function that, for a given input $\vec{x} \in \mathbb{R}^N$, is defined as

\begin{equation}
    \textrm{softmax}(\vec{x})_i = \frac{e^{x_i}}{\sum_{k=1}^N e^{x_k}}
\end{equation}
and  is ubiquitously used in deep learning systems to normalize raw model outputs into probabilities. By applying this function to the rows of $\Pi$ we effectively ensure that the resulting $\Pi$ satisfies the constraints in Eq. \ref{diag_op_prob}. 
Using this function makes the optimization problem non-convex, introducing the possibility of converging to local minima and saddle points. Nevertheless, gradient descent methods, especially when combined with stochastic minibatching, momentum, and adaptive learning rates, have been empirically successful at finding solutions of high quality in this regime \cite{kingma2015adam, panageas2019first}.
Note that other normalization techniques such as squaring normalization can also be used in place of the softmax function.
During a given iteration of the gradient descent algorithm the POVM parameters $\Pi$ are updated using the standard gradient descent update rule
\begin{align}
    \Pi' = \Pi - \gamma \nabla_\Pi\mathcal{L}(\Pi, F, P)
\end{align}
where $\gamma$ is the step size or learning rate and $\nabla_\Pi\mathcal{L}(\Pi, F, P)$ is the Euclidean gradient of the loss function from Eq. \ref{diag_op_prob}, $\mathcal{L}(\Pi, F, P) = || P - F\Pi ||_F^2$, with respect to $\Pi$.

Although we primarily focus on the tomography of detectors used to probe photon number we note that, given the formulation above, this method can be utilized to perform QDT for detectors whose POVM elements are diagonal in any particular orthonormal basis, not just the Fock basis, and that a broad set of detectors of interest satisfy this criteria. 
For continuous variable (qumode-based) systems notable examples such as bucket (light/no light) detectors along with photon number resolving detectors have POVM elements which are diagonal in the Fock basis and homodyne detectors have POVM elements which are diagonal in the quadrature basis. 
The latter observation was used in the experimental reconstruction of the POVM of a homodyne detector using a least squares method~\cite{grandi2017experimental}. 
For discrete variable (qubit-based) systems, which are particularly relevant for quantum computation, Pauli-$Z$ projective measurements are diagonal in the computational basis and the POVM elements describing a noisy multi-qubit measurement are prevalently assumed to be diagonal in the computational basis for qubit readout error models~\cite{hamilton2020scalable, bravyi2021mitigating,  geller2020rigorous, jattana2020general, nation2021scalable, malik2025coherence}.

\begin{figure*}[t]
\centering
\begin{minipage}{0.95\textwidth}
\begin{algorithm}[H]
\caption{Quantum detector tomography using Adam algorithm with learning rate decay}
\label{alg:adam}
\begin{algorithmic}[1]
\Require Initial parameters $\Pi_0$, 
initial learning rate $\gamma_0$, 
maximum number of iterations $T$, exponential decay rates $\beta_1, \beta_2 \in [0,1)$, 
batch size $B$, 
experimental data $P$,
probe states $F$,
learning rate decay rate $\chi$,
small constant $\epsilon$ for numerical stability (typically set to $10^{-8}$) 
\State $m_0 \leftarrow 0$ \Comment{Initialize first moment variable}
\State $v_0 \leftarrow 0$ \Comment{Initialize second moment variable}
\State $t \leftarrow 0$ \Comment{Current iteration number (epoch number)}
\While{$t< T$}
    \State $t \leftarrow t + 1$
    \While{all probe states in $F$ not yet sampled} \Comment{An epoch is one pass over all data points}
    \State Sample batch of $B$ probe states, $F_B$, from $F$  and corresponding data, $P_B$, from $P$
    \State Apply softmax function along rows of  $\Pi_{t-1}$ \Comment{Enforce POVM constraints}
    \State $g_t \leftarrow \nabla_{\Pi_{t-1}} \mathcal{L}(\Pi_{t-1}, F_B, P_B)$ \Comment{Compute Euclidean gradient for batch}
    \State $m_t \leftarrow \beta_1 m_{t-1} + (1-\beta_1) g_t$ \Comment{Update biased first moment estimate}
    \State $v_t \leftarrow \beta_2 v_{t-1} + (1-\beta_2) g_t \odot g_t$ \Comment{Update biased second moment estimate. $\odot$ denotes Hadamard product.}
    \State $\hat{m}_t \leftarrow \dfrac{m_t}{1-\beta_1^t}$ \Comment{Correct bias in first moment estimate}
    \State $\hat{v}_t \leftarrow \dfrac{v_t}{1-\beta_2^t}$ \Comment{Correct bias in second moment estimate}
    \State $\Pi_t \leftarrow \Pi_{t-1} - 
        \gamma \dfrac{\hat{m}_t}{\sqrt{\hat{v}_t} + \epsilon}$ \Comment{Compute parameter update}
    \EndWhile
    \State $\gamma_t \leftarrow \gamma_{0}\chi^{t}$ \Comment{Decay learning rate on exponential schedule}
\EndWhile 
\State Apply softmax function along rows of  $\Pi_{t}$ \Comment{Enforce POVM constraints} \\
\Return $\Pi_t$
\end{algorithmic}
\end{algorithm}
\end{minipage}
\end{figure*}

\section{Numerical Experiments}{\label{num_exp}}
We benchmark our approach against QDT based on CCO in numerical experiments in which we characterize the POVM of photon number resolving detectors 
and randomly generated mulit-qubit measurement POVMs which are diagonal in the computational basis. 
For our coherent state probes we use $D$ evenly spaced values of the mean photon number $|\alpha|^2$ between $[0,|\alpha|^2_{\textrm{max}}]$ where $|\alpha|_{\textrm{max}}^2$ is chosen to be the largest value smaller than $M$ that satisfies 
\begin{align}
    1-\Phi(M-1; |\alpha|^2_{\textrm{max}}) \leq 10^{-5}
\end{align}
where $\Phi(M-1; |\alpha|^2_{\textrm{max}})$ is the cumulative distribution function of the Possion distribution in Eq. \ref{Poisson_eq} with $j=M-1$. This ensures our probes span the full Hilbert space without having support greater than $10^{-5}$ outside our truncation.
We keep the phase of the probe states fixed as we are considering phase insensitive detectors. 
For general $n$-qubit detector tomography the set of tomographically complete states is given by 
\begin{align}
    \left\{
    \begin{bmatrix} 1 \\ 0 \end{bmatrix}, \begin{bmatrix} 0 \\ 1 \end{bmatrix},
    \frac{1}{\sqrt{2}}\begin{bmatrix} 1 \\ 1 \end{bmatrix},
    \frac{1}{\sqrt{2}} \begin{bmatrix} 1 \\ i \end{bmatrix} \right\}^{\otimes n}
\end{align}
which leads to the number of probes scaling as $4^n$.
However, under the assumption that the detector POVM is diagonal in the computational basis the set of tomographically complete probe states is given by the
set \cite{hamilton2020scalable}
% hamilton2020scalable,jattana2020general explcitly say they use 2^n probe states which means they only consider the diagonal POVM case
\begin{align}\label{eq:diag_probes_dv}
    \left\{
    \begin{bmatrix} 1 \\ 0 \end{bmatrix}, \begin{bmatrix} 0 \\ 1 \end{bmatrix} \right\}^{\otimes n}.
\end{align}
which scales as $2^n$.
Similar to the continuous variable case the probe state matrix $F$ is now given by
\begin{align}
    F_{ij} = |\psi_{ij}|^2
\end{align}
where $|\psi_{ij}|^2$ is the squared modulus of the $j$-th ket element of probe state $i$ 
or, equivalently, the the $j$-th diagonal element of the density matrix of probe state $i$.  
Following Refs. \cite{lundeen2009tomography, zhang2012recursive, zhang2012mapping, schapeler2024scalable} we calculate the fidelity $\mathcal{F}(\hat{E}_n, E_n)$ between the reconstructed POVM element $\hat{E}_n$ and the true POVM element $E_n$ as

\begin{align}
    \mathcal{F}(\hat{E}_n, E_n) =\frac{\textrm{Tr}^2\left[\sqrt{\sqrt{E_n}\hat{E}_n\sqrt{E_n}}\right]}{\textrm{Tr}[\hat{E}_n]\textrm{Tr}[E_n]}
\end{align}
and subsequently define the average reconstruction fidelity 
\begin{align}
\bar{\mathcal{F}} = \frac{\sum_{i=1}^N\mathcal{F}(\hat{E}_i, E_i)} {N}
\end{align} 
as the average fidelity across all elements in the POVM.
We also define the mean squared error (MSE) between the reconstructed POVM element $\hat{E}_n$ and the true POVM element $E_n$ as
\begin{align}
\textrm{MSE}(\hat{E}_n, E_n) = \frac{1}{M}||\hat{E}_n - E_n||_F^2 
\end{align}
and the average MSE as 
\begin{align}
\frac{\sum_{i=1}^N\textrm{MSE}(\hat{E}_i, E_i)} {N}.
\end{align}

For our particular gradient descent algorithm we use the Adaptive Moment Estimation (Adam) algorithm implemented in \texttt{PyTorch} \cite{paszke2019pytorch} which falls under the broader class of stochastic gradient descent algorithms which derive their name from the fact that they compute parameter updates using random subsets (minibatches) of the data \cite{goodfellow2016deep}.
The pseudo-code for the Adam algorithm is provided in Algorithm \ref{alg:adam} with further details available in~\cite{goodfellow2016deep, bishop2023deep}. 
We set a maximum number of iterations of $T=100$ as our stopping criterion.
Stochastic gradient descent algorithms are more memory efficient than batch gradient descent methods, which use the whole dataset to compute parameter updates, and allow for the possibility of escaping from local minima \cite{bishop2023deep} which is particularity useful in our case as our cost function is non-convex.  
We use a learning rate scheduler that decays the learning rate exponentially after each iteration of gradient descent.
The hyperparameters we used for the scheduler and Adam algorithm in our experiments are shown in Table \ref{tab:hyperparam_table} and could be further optimized via additional hyperparameter tuning.
The values in Table \ref{tab:hyperparam_table} were informed by a hyperparameter sweep using the hyperparameter tuning framework \texttt{Optuna}~\cite{akiba2019optuna}.
Importantly, these values were obtained by optimizing the loss function and not the reconstruction fidelity or the MSE between the true and reconstructed POVMs so as to adhere to machine learning best practices.

We benchmark our method against the state-of-the-art CCO solver MOSEK, through \texttt{CVXPY} \cite{diamond2016cvxpy}, which implements a primal-dual interior-point method. We run all experiments on a desktop with an AMD Ryzen 9 3900X 12-Core Processor and 32 GB of RAM but note that our approach can seamlessly be adapted to utilize GPUs to further accelerate optimization speed and scale to larger problem sizes.  

\begin{table}[ht]
\centering
\begin{tabular}{|c|c|}
\hline
Hyperparameter & Value \\
\hline
Learning rate ($\gamma$) & $0.01$ \\
\hline
Learning rate decay ($\chi$) & $0.999$ \\
\hline
 $1$st moment estimates decay rate ($\beta_1$) & $0.9$ \\
\hline
$2$nd moment estimates decay rate ($\beta_2$) & $0.9$ \\
\hline
Number of iterations (epochs) & $100$ \\
\hline
Batch size & $25$ \\
\hline
\end{tabular}
\caption{Hyperparameters used when running the gradient descent algorithm. }
\label{tab:hyperparam_table}
\end{table}

\subsection{Photon number resolving detectors}
In quantum optics the prototypical example of a measurement whose POVM elements are diagonal in the Fock basis is the one performed by photon number resolving (PNR) detectors \cite{kok2010introduction}. These detectors have recently reached technological maturity in the form of superconducting transition edge sensors (TESs) \cite{Lita2008} and superconducting nanowire single-photon detectors (SNSPDs) \cite{Cahall2017}. 
The former has been used to count up to 100 photons for the use of quantum random-number-generation \cite{Eaton2023} and the latter  in principle allows measurement of on the order of $10^5$ photons per input pulse via time-multiplexing \cite{tiedau2019high}. 
TESs have also been used in heralding schemes to prepare optical Gottesman–Kitaev–Preskill states \cite{larsen2025integrated} which are a key resource for universal fault-tolerant photonic quantum computing \cite{Gottesman2001, Baragiola2019}. 

For an ideal PNR detector which can resolve up to $N-1$ photons the POVM is given by the following set \cite{feito2009measuring}

\begin{align}
    \left\{\ket{0}\bra{0}, \dots, \ket{N-1}\bra{N-1}, \mathbb{I}_M - \sum_{k=0}^{N-1}\ket{k}\bra{k}\right\}.
\end{align}
The POVM element $E_n$ for a PNR detector with quantum efficiency $0 \leq \eta \leq 1$ is

\begin{align}
    E_n =\sum_{k=0}^{M-1} \binom{k}{n} \eta^n(1-\eta)^{k-n}\ket{k}\bra{k}
\end{align}
where $M$ is the Hilbert space dimension and the POVM is given by

\begin{align}
    \left\{E_0,...,E_{N-1}, \mathbb{I}_M - \sum_{k=0}^{N-1}E_k\right\}.
\end{align}
For realistic detectors such as those that have quantum efficiency below unity a next-neighbor regularization term is often added to the objective function in Eq. \ref{diag_op_prob}

\begin{equation}\label{diag_op_prob_reg}
\begin{aligned}
    \min_{\Pi} \quad  & || P - F\Pi ||_F^2 + \lambda\sum_{j=1}^{N} \sum_{i=1}^{M-1}(\Pi_{i,j} - \Pi_{i+1,j})^2\\
\textrm{subject to} \quad & \Pi_{ij} \geq 0 \\
  &\sum_{j=1}^{N}\Pi_{ij} = 1 \quad \forall i \in \{1, \cdots, N\}
\end{aligned}
\end{equation}
which is motivated by the fact that for such detectors the elements on the diagonal of $E_j$ should vary smoothly \cite{feito2009measuring}. This is part of a broader class of regularization techniques called smoothing regularization \cite{boyd2004convex}. The parameter $\lambda \geq 0$ is used to control the strength of this regularization penalty. 

For our experiments we set $N=25$ leading to a detector that can resolve up to $24$ photons.  We use $D=2000$ coherent state probes and a Hilbert space of size $M=200$. We test our QDT method by reconstructing the POVM of an ideal PNR detector and one with quantum efficiency $\eta=0.85$ which is within the efficiency range of existing PNR detectors \cite{fukuda2011titanium, zhang2017nbn, schapeler2021quantum, Eaton2023}. % 85% was the effiency of the high dynamic range SNSPD used in schapeler2021quantum
When reconstructing the POVM of an ideal detector we set the regularization parameter $\lambda$ to $0$ and set it to $10^{-5}$ in the inefficient case. Due to the stochastic nature of our gradient descent algorithm we run the algorithm 20 times with different random initializations for the POVM parameters and plot the mean and standard deviation of the results when presenting plots of average reconstruction fidelity.

\subsubsection{Time complexity}
\begin{figure}
    \centering
    \begin{subfigure}[b]{0.475\textwidth}
    \centering
    \includegraphics[width=1.025\columnwidth]{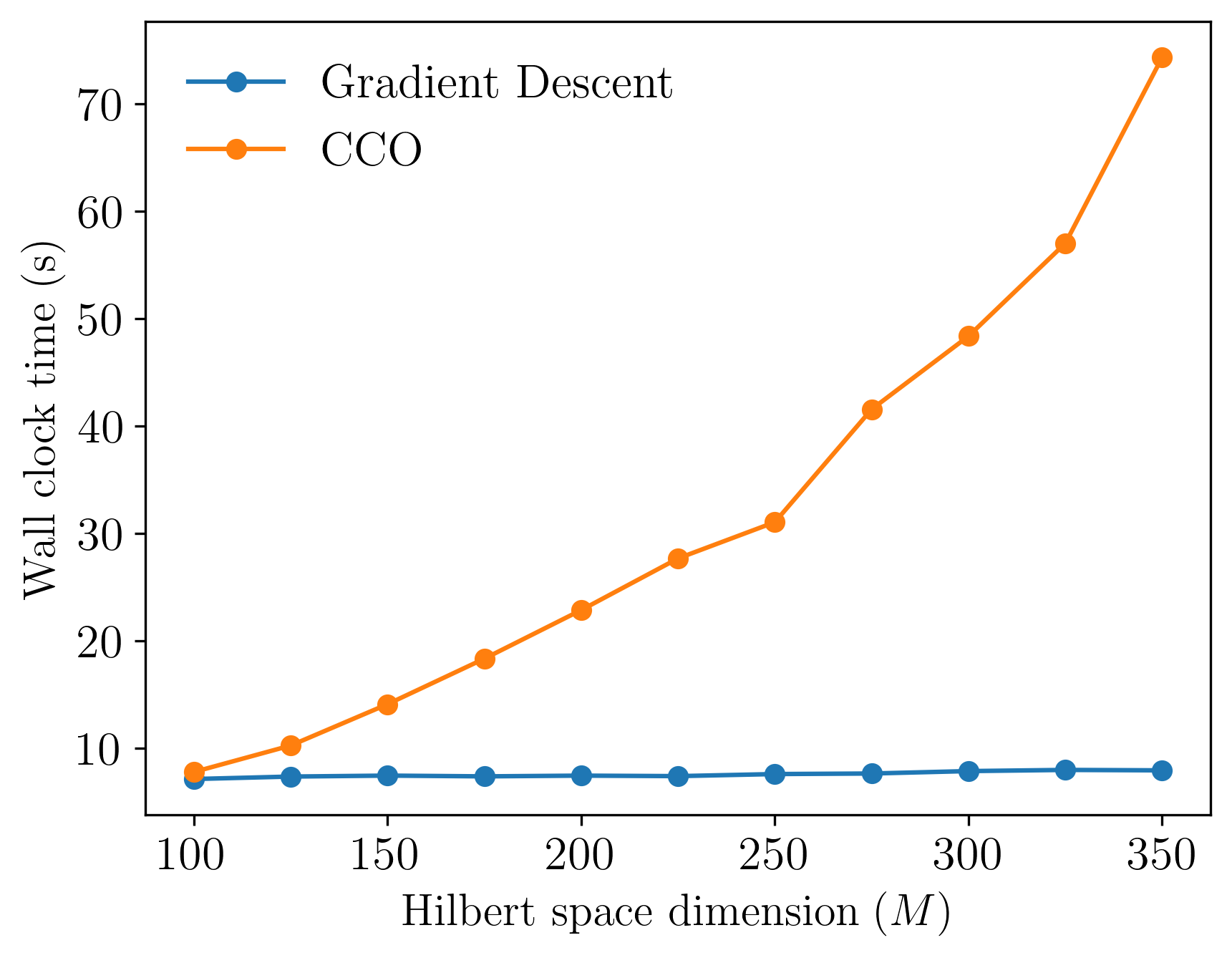}
    \caption{}
    \label{fig:ideal_wall_clock}
    \end{subfigure}
    \begin{subfigure}[b]{0.475\textwidth}
    \centering
    \includegraphics[width=1.025\columnwidth]{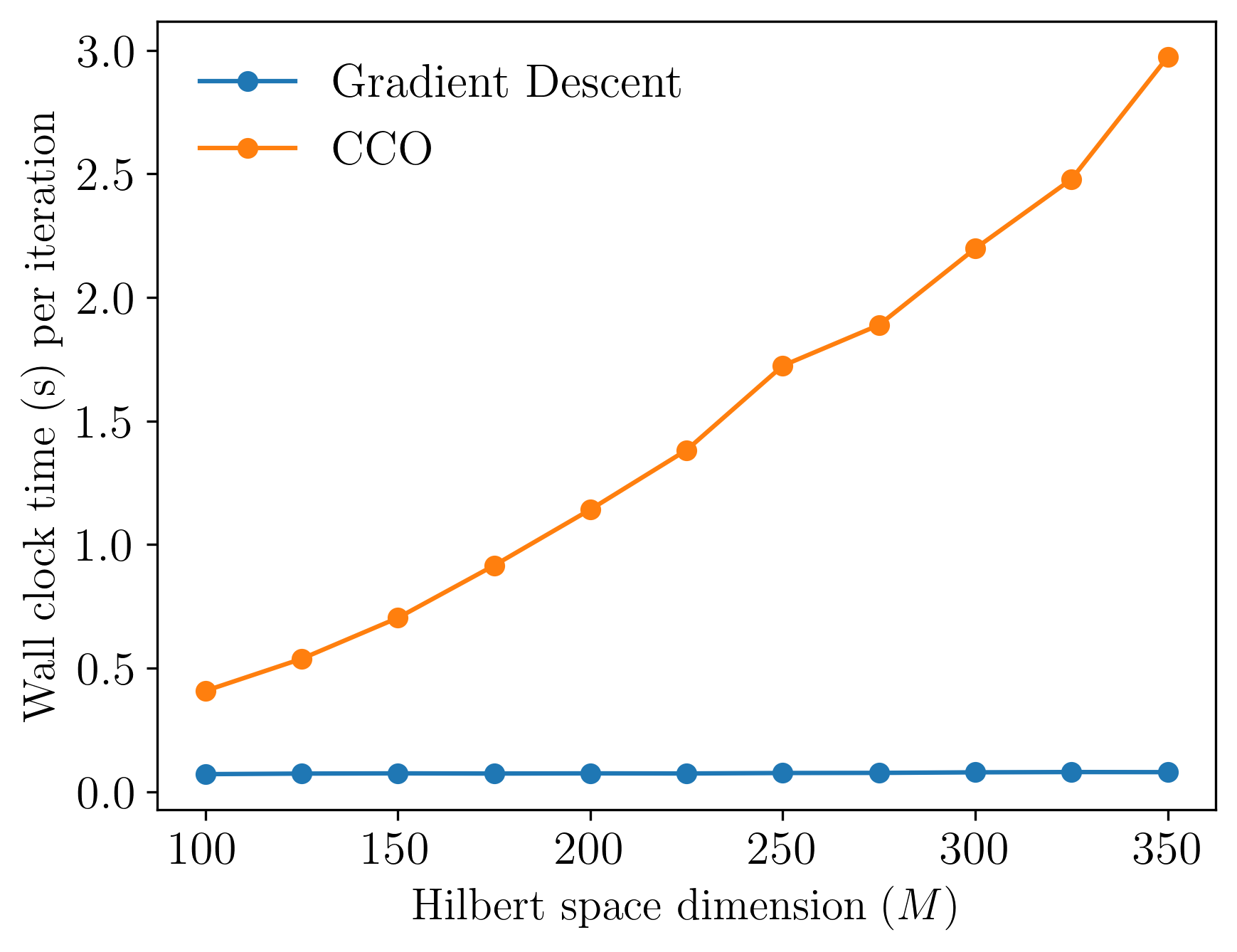}
    \caption{}
     \label{fig:ideal_wall_iter}
    \end{subfigure}
    \caption{Wall clock time (in seconds) and wall clock time per iteration for QDT methods as Hilbert space size increases.}
    \label{fig:time_comp}
\end{figure}

PNR detectors, such as those implemented as SNSPDs, can exhibit sensitivity to Hilbert spaces of very large dimension e.g. on the order of $10^5$ or more~\cite{tiedau2019high, schapeler2021quantum}, via use of time-multiplexed detector design \cite{Fitch2003}.  Therefore, it is important to examine how the time complexity of a QDT approach scales with the Hilbert space dimension.
Fig. \ref{fig:ideal_wall_clock} shows the wall clock time taken for each algorithm when optimizing Eq. \ref{diag_op_prob_reg} for an ideal PNR detector. 
To control for the fact that the gradient descent algorithm is always run with a fix number of iterations in our experiments Fig. \ref{fig:ideal_wall_iter} shows the wall clock time taken per iteration for each algorithm. 
The plot shows that the time per iteration for CCO follows a similar trend as the overall wall clock time indicating that as the problem size increases the cost for each iteration of the CCO method greatly increases. The gradient descent method by contrast remains relatively constant in both its wall clock time and per iteration wall clock time. 
This is due to the fact that, during each iteration, interior point CCO methods compute both gradients and the Hessian matrix and then must solve the Karush–Kuhn–Tucker (KKT) system of linear equations which depend on these objects \cite{boyd2004convex}. 
This is much more expensive compared to gradient descent which only needs to compute the gradients of the cost function with respect to the free variables. 

\begin{figure}
    \centering
    \begin{subfigure}[b]{0.475\textwidth}
    \centering
    \includegraphics[width=1.025\columnwidth]{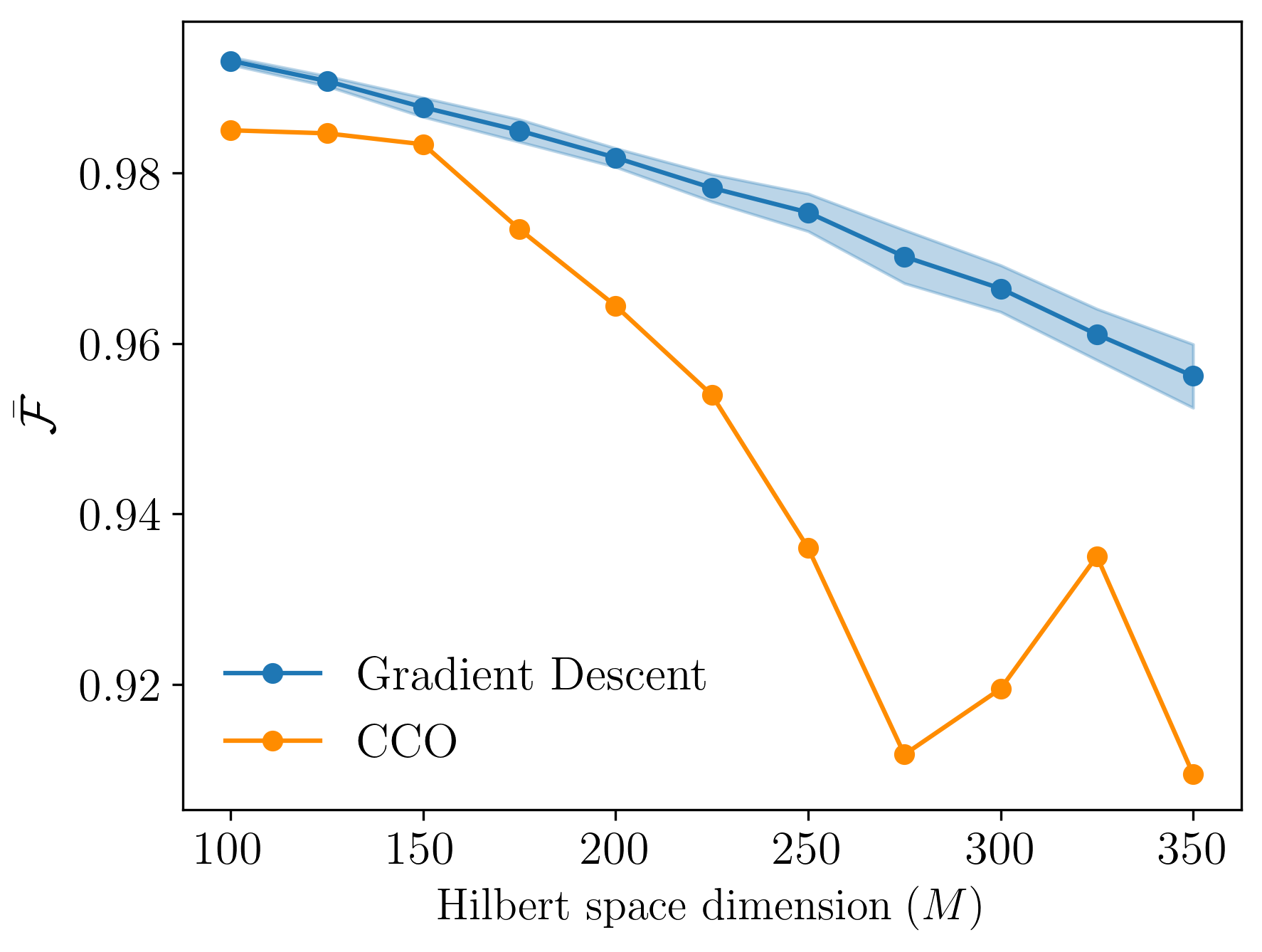}
    \caption{}
     \label{fig:ideal_fidelity}
    \end{subfigure}
    \begin{subfigure}[b]{0.475\textwidth}
    \centering
    \includegraphics[width=1.025\columnwidth]{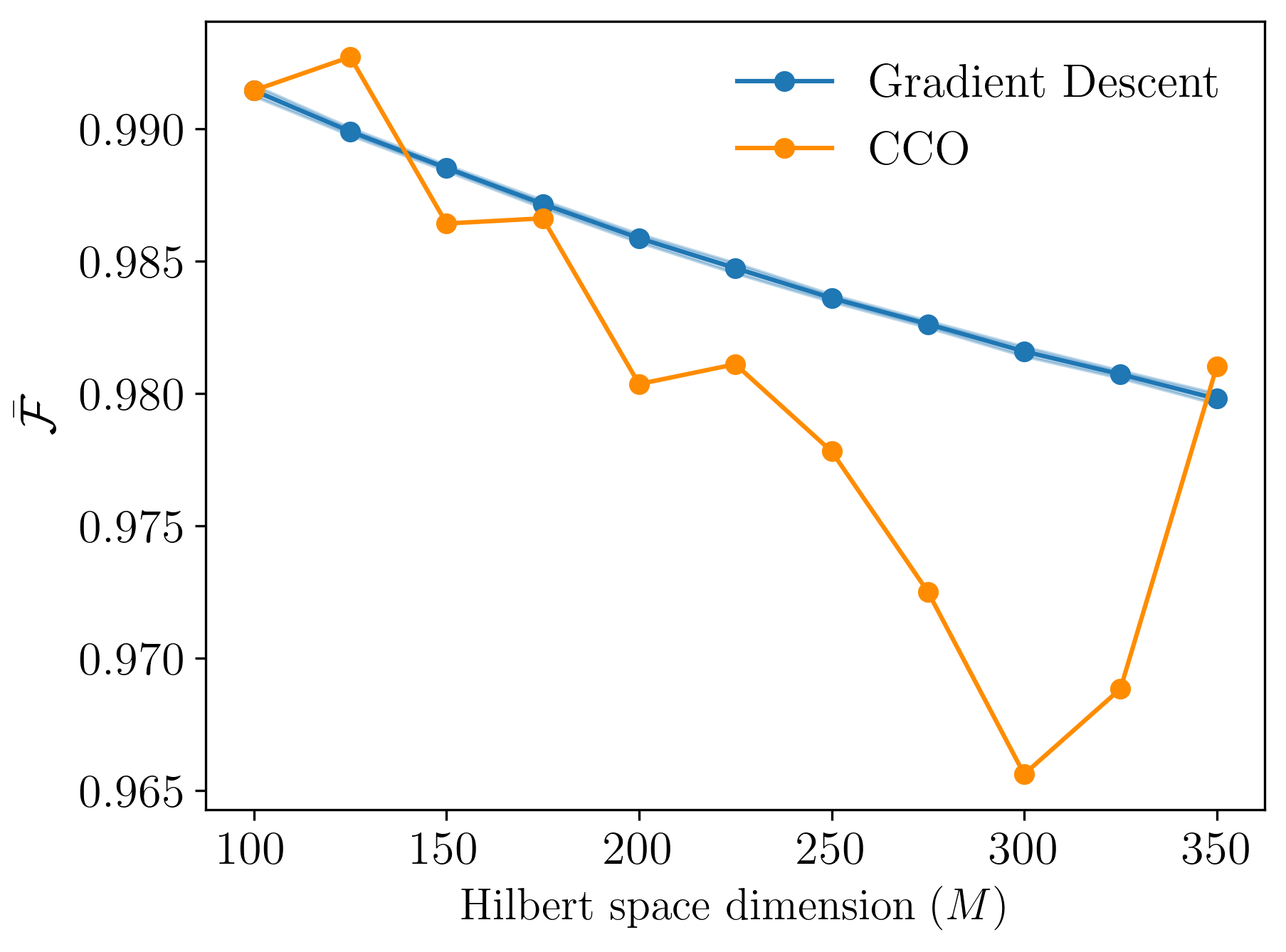}
    \caption{}
     \label{fig:lossy_fidelity}
    \end{subfigure}
    \caption{Average reconstruction fidelity of QDT methods when using different Hilbert space truncations. Gradient descent curve shows mean value of $\bar{\mathcal{F}}$ over 20 different runs of the algorithm each with different random parameter initialization. Shaded region shows one standard deviation. (a) Average POVM reconstruction fidelity for ideal PNR detector. (b) Average POVM reconstruction fidelity for PNR detector with $85\%$ efficiency. }
    \label{fig:time_comp_fidelity}
\end{figure}

\begin{figure}
    \centering
    \begin{subfigure}[b]{0.475\textwidth}
    \centering
    \includegraphics[width=1.025\columnwidth]{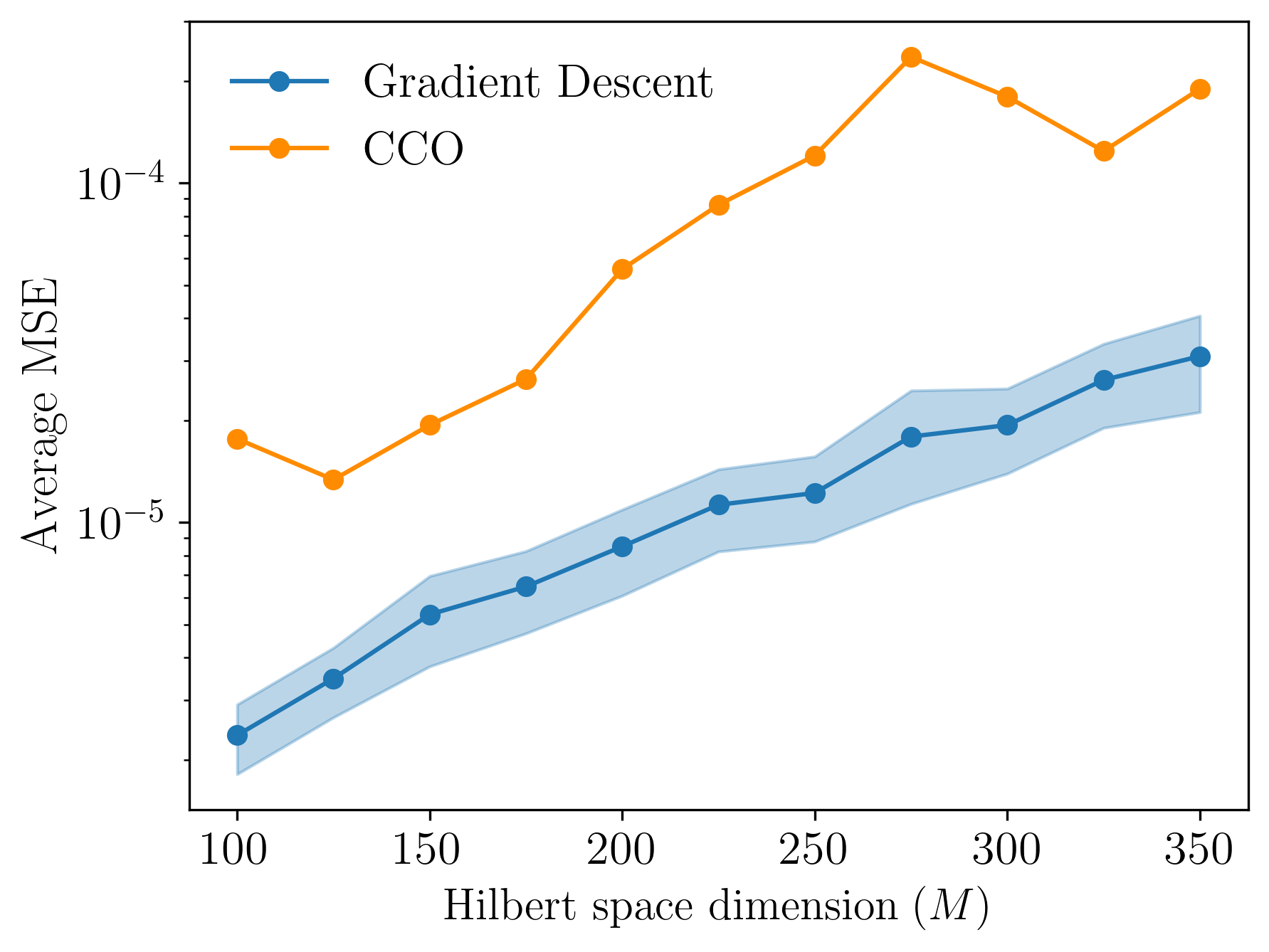}
    \caption{}
     \label{fig:ideal_mse}
    \end{subfigure}
    \begin{subfigure}[b]{0.475\textwidth}
    \centering
    \includegraphics[width=1.025\columnwidth]{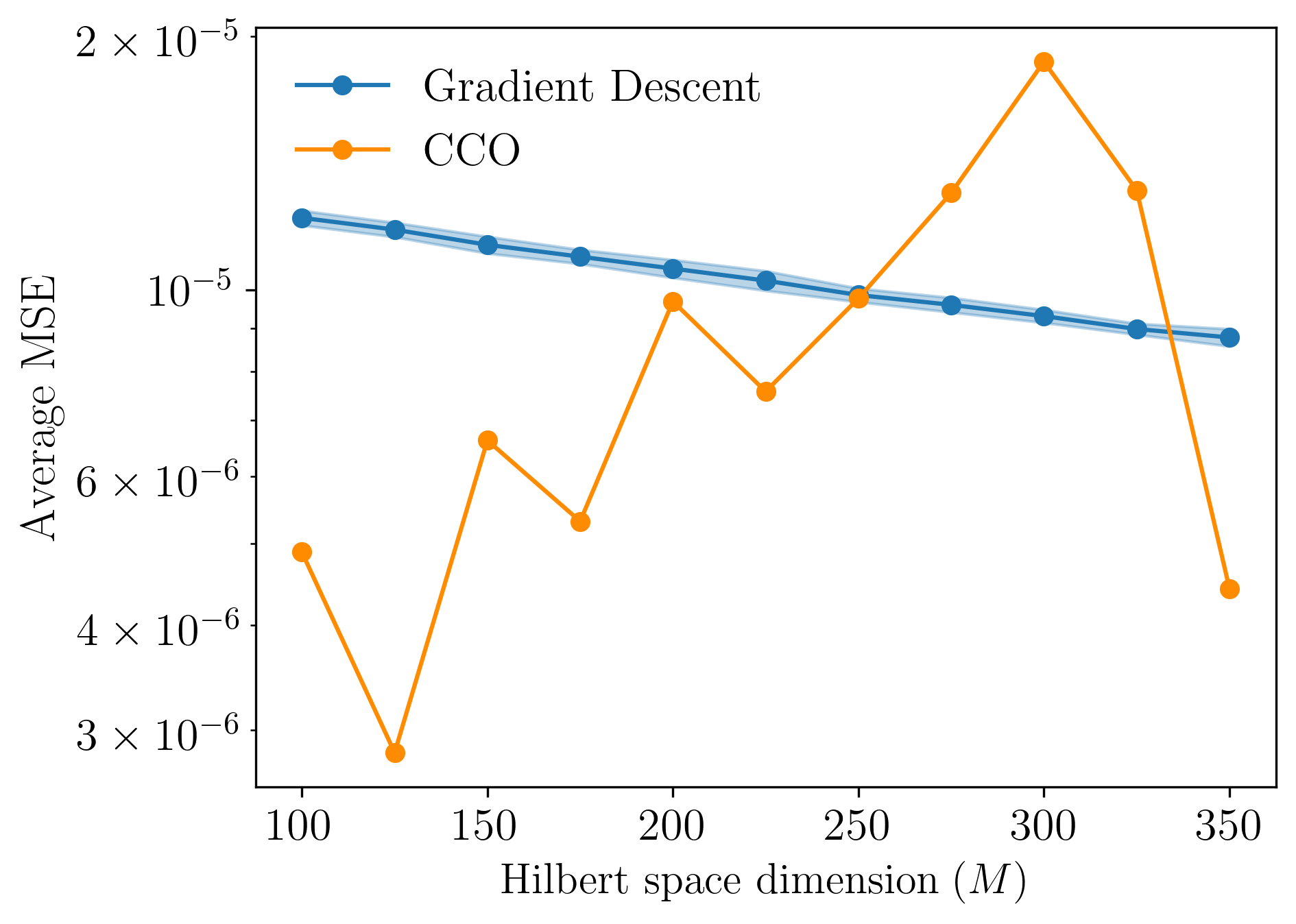}
    \caption{}
     \label{fig:lossy_mse}
    \end{subfigure}
    \caption{Average MSE of QDT methods when using different Hilbert space truncations. Gradient descent curve shows mean value of MSE over 20 different runs of the algorithm each with different random parameter initialization. Shaded region shows one standard deviation. (a) Average POVM MSE for ideal PNR detector. (b) Average POVM MSE for PNR detector with $85\%$ efficiency.} 
    \label{fig:time_comp_mse}
\end{figure}

Fig. \ref{fig:time_comp_fidelity} shows the average reconstruction fidelity of the two methods as the Hilbert space truncation is increased for both an ideal and $85\%$ efficient PNR detector. We can see that the gradient descent algorithm achieves higher values of $\bar{\mathcal{F}}$ in most cases. However, we note the average fidelity degrades for both methods as the Hilbert space dimension increases. This is likely because we do not increase the number of probe states used as we increase the Hilbert space dimension which increase the number of free variables in our problem without increasing the dataset size. 
We see that similar trends can be observed in the average MSE plots in Fig. \ref{fig:time_comp_mse}.

\subsubsection{Memory complexity}
Previous QDT approaches perform tomography by CCO utilizing interior-point methods \cite{lundeen2009tomography, schapeler2021quantum, liu2023optimized}, with the MOSEK solver serving as an example. 
As mentioned previously these algorithms require second-order information about the objective function that necessitates computation of the Hessian matrix \cite{bertsekas1982projected, boyd2004convex}. 
Due to this Ref. \cite{liu2023optimized} argues that in practice the main obstacle in detector tomography is limited memory resources.
In contrast gradient descent only requires computation and storage of the gradients which requires memory that scales linearly with the number of free variables $MN$ unlike the Hessian matrix whose computationally memory requirement is quadratic in the number of free variables \cite{bishop2023deep}. 
Stochastic gradient descent can further be made more memory efficient by use of gradient accumulation techniques which allow one to use a smaller minibatch size, therefore reducing the memory required for an iteration, while still mimicking one of the effects of using a larger minibacth size i.e. obtaining a less noisy estimate of the gradients.   
% The Adam algorithm requires more memory than vanilla gradient descent due to storing the gradients as well as two momentum vectors computed from past gradients but this is a constant factor increase so it is still linear.

\subsubsection{Experimental noise}
\begin{figure}
    \centering
    \begin{subfigure}[b]{0.475\textwidth}
    \centering
    \includegraphics[width=1.025\columnwidth]{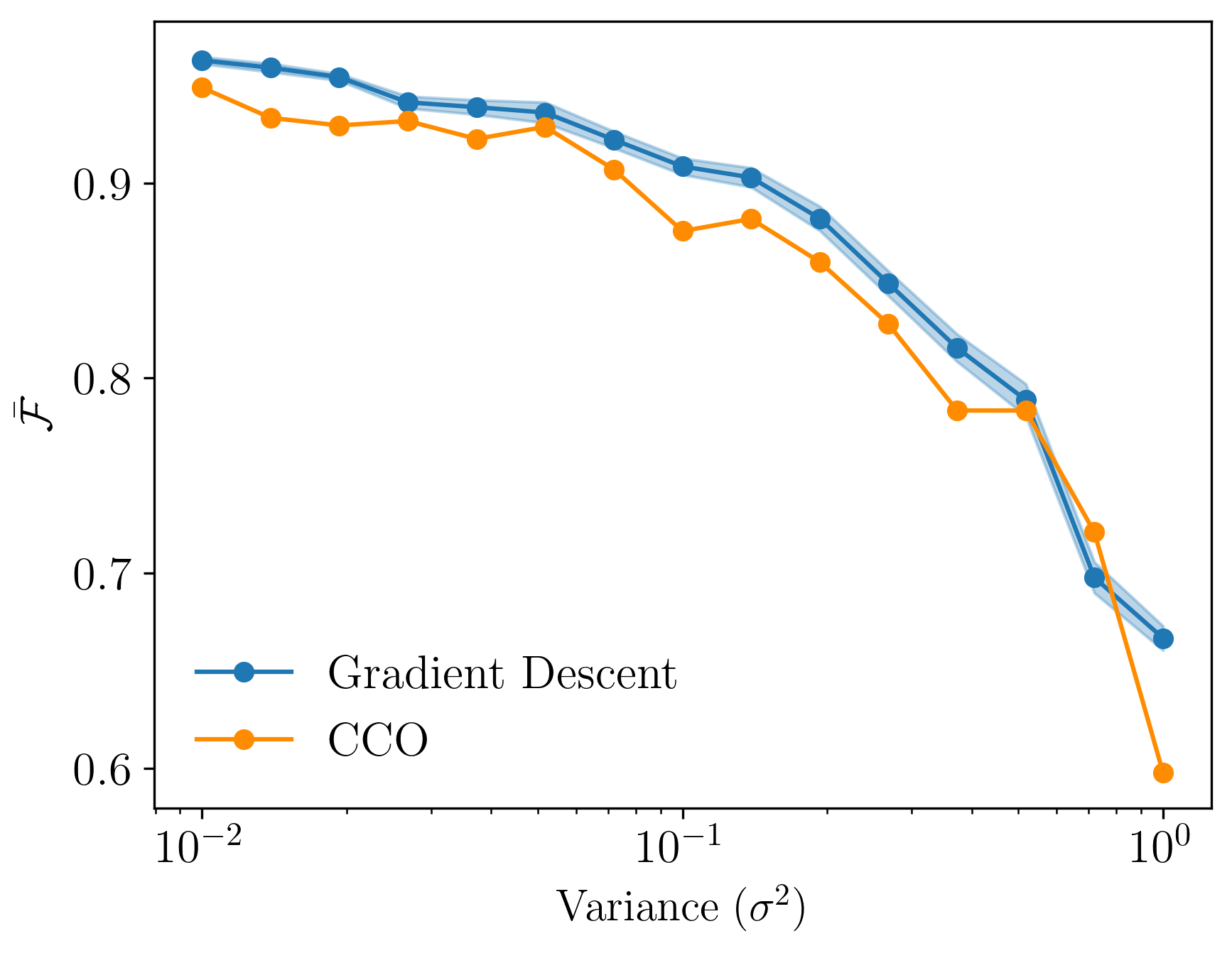}
    \caption{}
    \label{fig:ideal_noise}
    \end{subfigure}
    \begin{subfigure}[b]{0.475\textwidth}
    \centering
    \includegraphics[width=1.025\columnwidth]{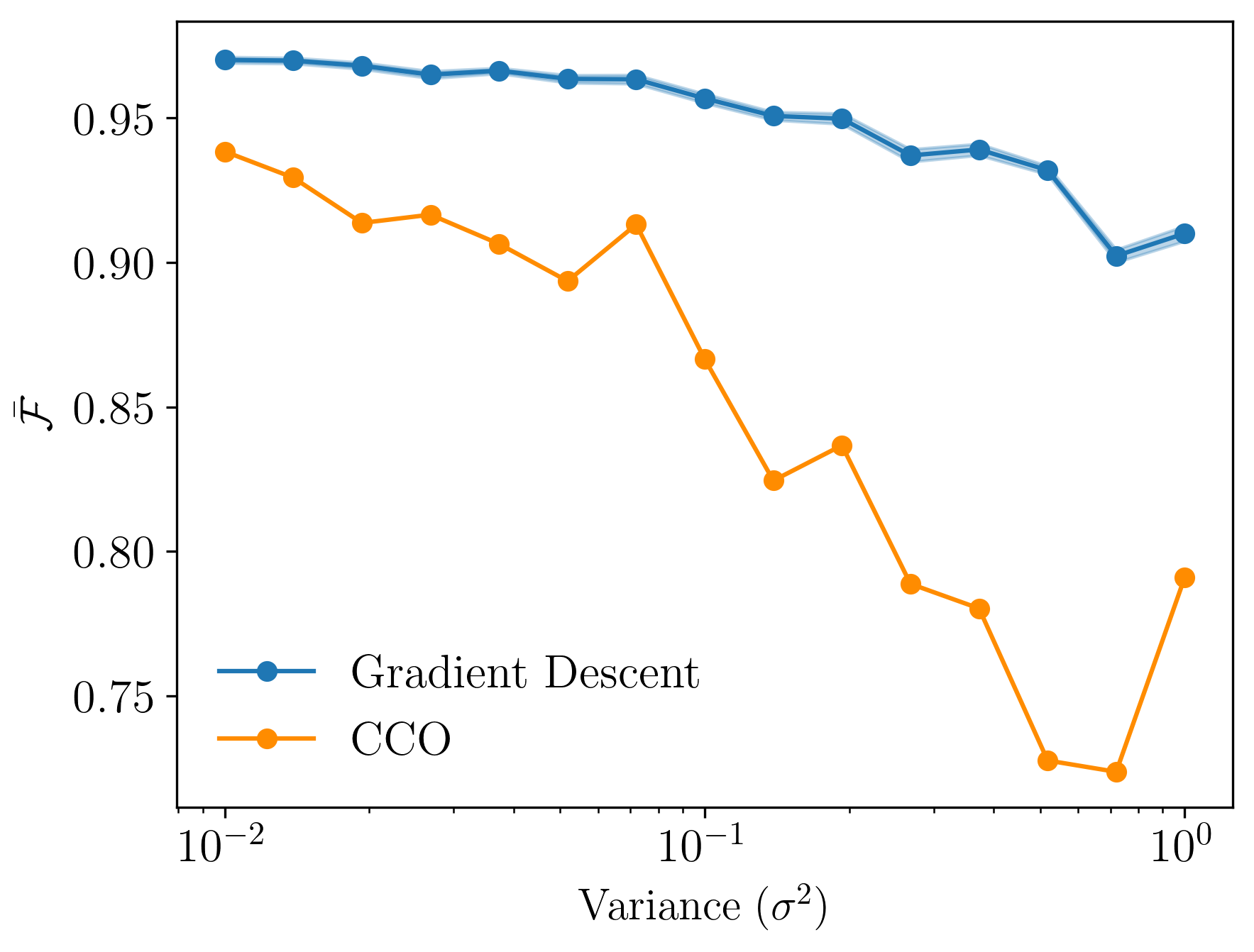}
    \caption{}
     \label{fig:lossy_noise}
    \end{subfigure}
    \caption{Average reconstruction fidelity of QDT methods under varying levels of Gaussian noise $\mathcal{N}(0, \sigma^2)$ in the coherent state amplitudes $|\alpha|^2$. (a) Average POVM reconstruction fidelity for ideal PNR detector. (b) Average POVM reconstruction fidelity for PNR detector with $85\%$ efficiency. }
    \label{fig:stat_noise}
\end{figure}

\begin{figure}
    \centering
    \begin{subfigure}[b]{0.475\textwidth}
    \centering
    \includegraphics[width=1.025\columnwidth]{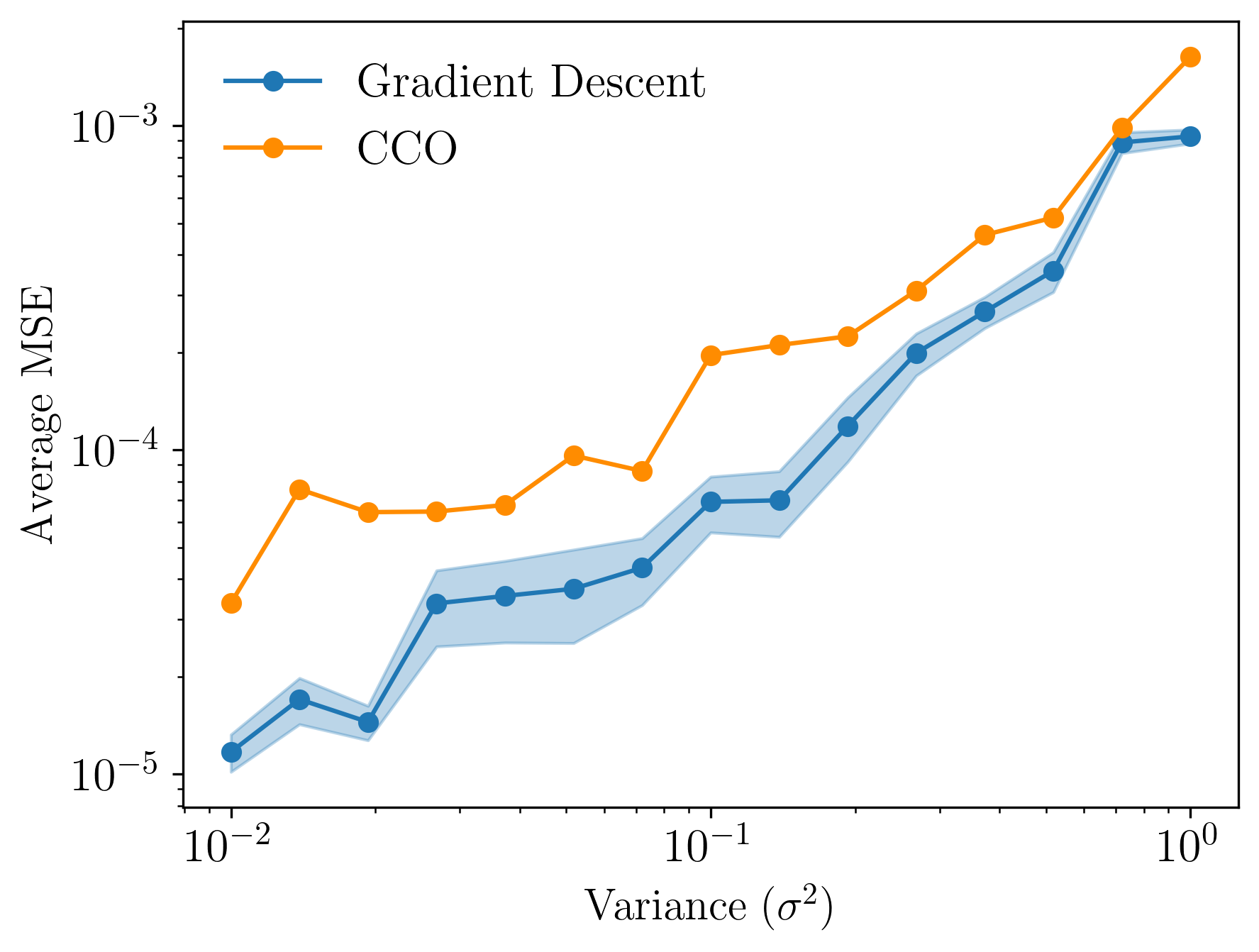}
    \caption{}
    \label{fig:ideal_noise_mse}
    \end{subfigure}
    \begin{subfigure}[b]{0.475\textwidth}
    \centering
    \includegraphics[width=1.025\columnwidth]{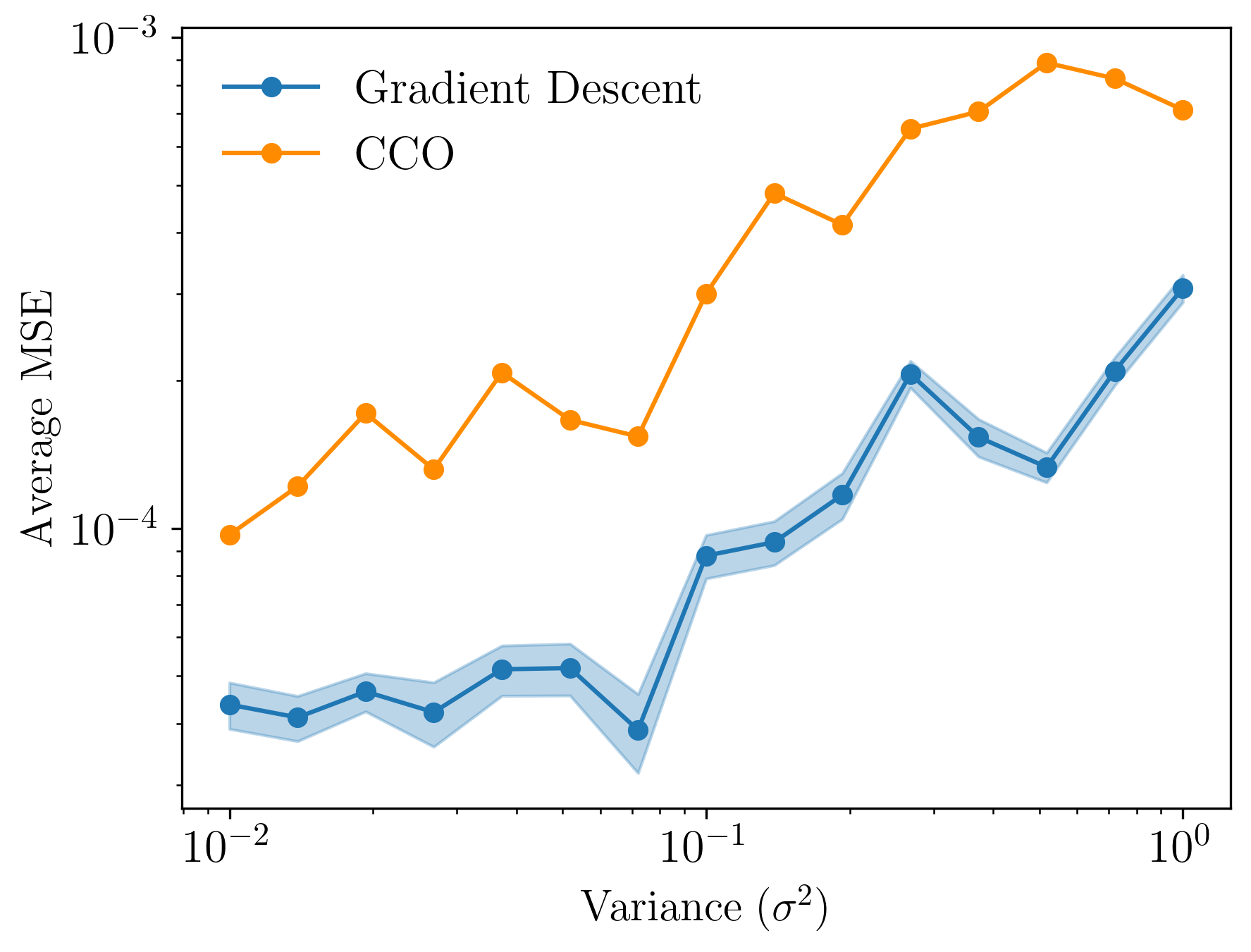}
    \caption{}
     \label{fig:lossy_noise_mse}
    \end{subfigure}
    \caption{Average MSE of QDT methods under varying levels of Gaussian noise $\mathcal{N}(0, \sigma^2)$ in the coherent state amplitudes $|\alpha|^2$. (a) Average POVM MSE for ideal PNR detector. (b) Average POVM MSE for PNR detector with $85\%$ efficiency. }
    \label{fig:stat_noise_mse}
\end{figure}

One source of errors in the QDT process is in mismatches between the probe states generated experimentally to construct $P$ and the assumed probe states used in $F$. Here we investigate the impact of this statistical noise in the probe states used to generate the experimental statistics in $P$. Specifically we examine uncertainty in the amplitude of the laser used to generate the coherent state probes. Following Ref. \cite{feito2009measuring} we simulate this by adding Gaussian noise with mean $0$ and variance $\sigma^2$ to the values of $|\alpha|^2$ used to compute the values of $P$. 
The results in Fig. \ref{fig:ideal_noise} show that for an ideal detector both algorithms achieve comparable reconstruction fidelity for all noise levels with gradient descent typically performing slightly better. 
In the case of an inefficient detector Fig. \ref{fig:lossy_noise} shows that gradient descent is more resilient to amplitude noise while CCO shows comparable performance to the ideal detector case.  
Fig. \ref{fig:stat_noise_mse} shows that the corresponding reconstruction MSEs exhibit the same trend as the reconstruction fidelity.
%Maybe also look at what happens when using lossy coherent states

\subsubsection{Limited data}
\begin{figure}
    \centering
    \begin{subfigure}[b]{0.475\textwidth}
    \centering
    \includegraphics[width=1.025\columnwidth]{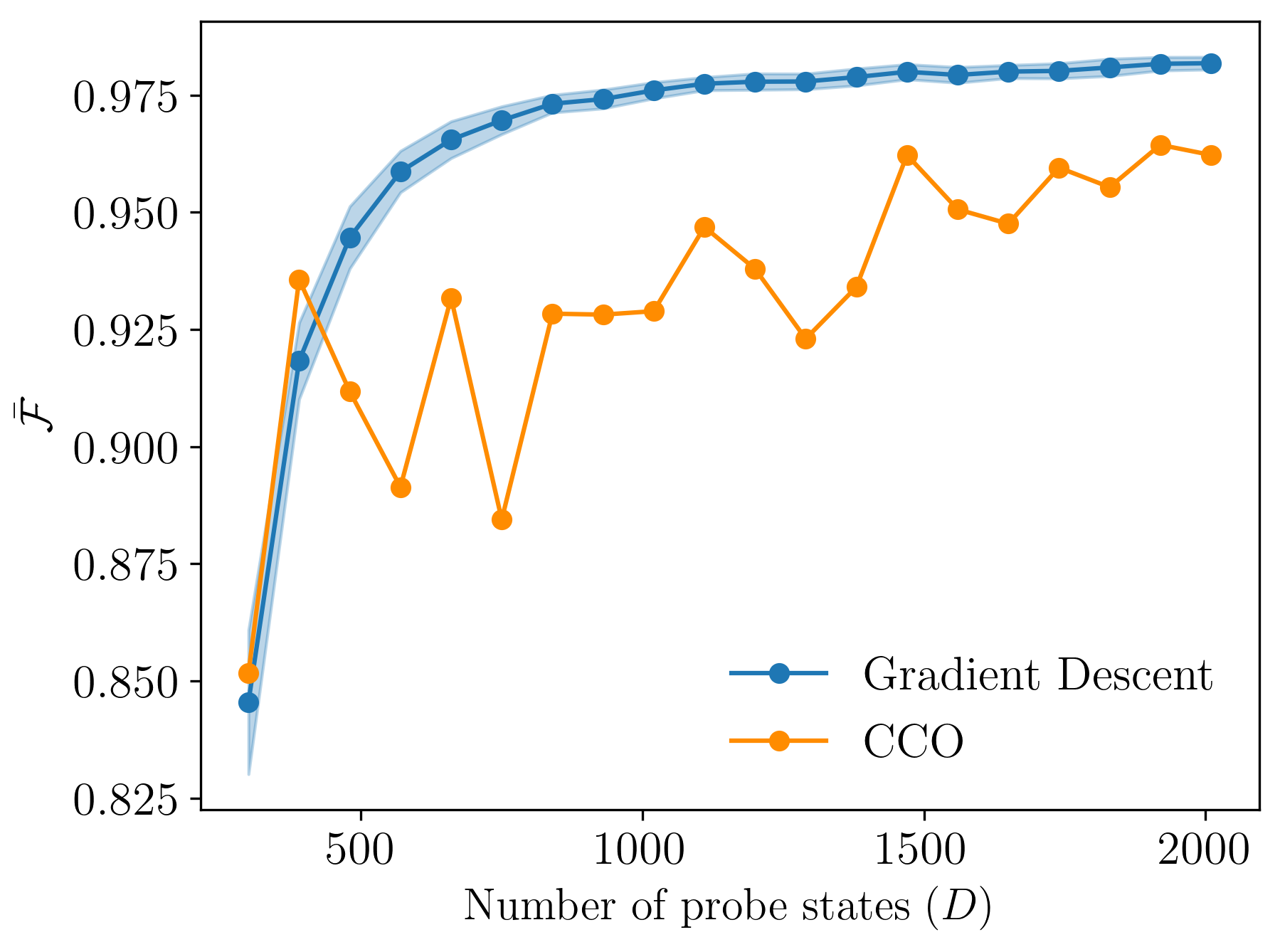}
    \caption{}
    \label{fig:ideal_data}
    \end{subfigure}
    \begin{subfigure}[b]{0.475\textwidth}
    \centering
    \includegraphics[width=1.025\columnwidth]{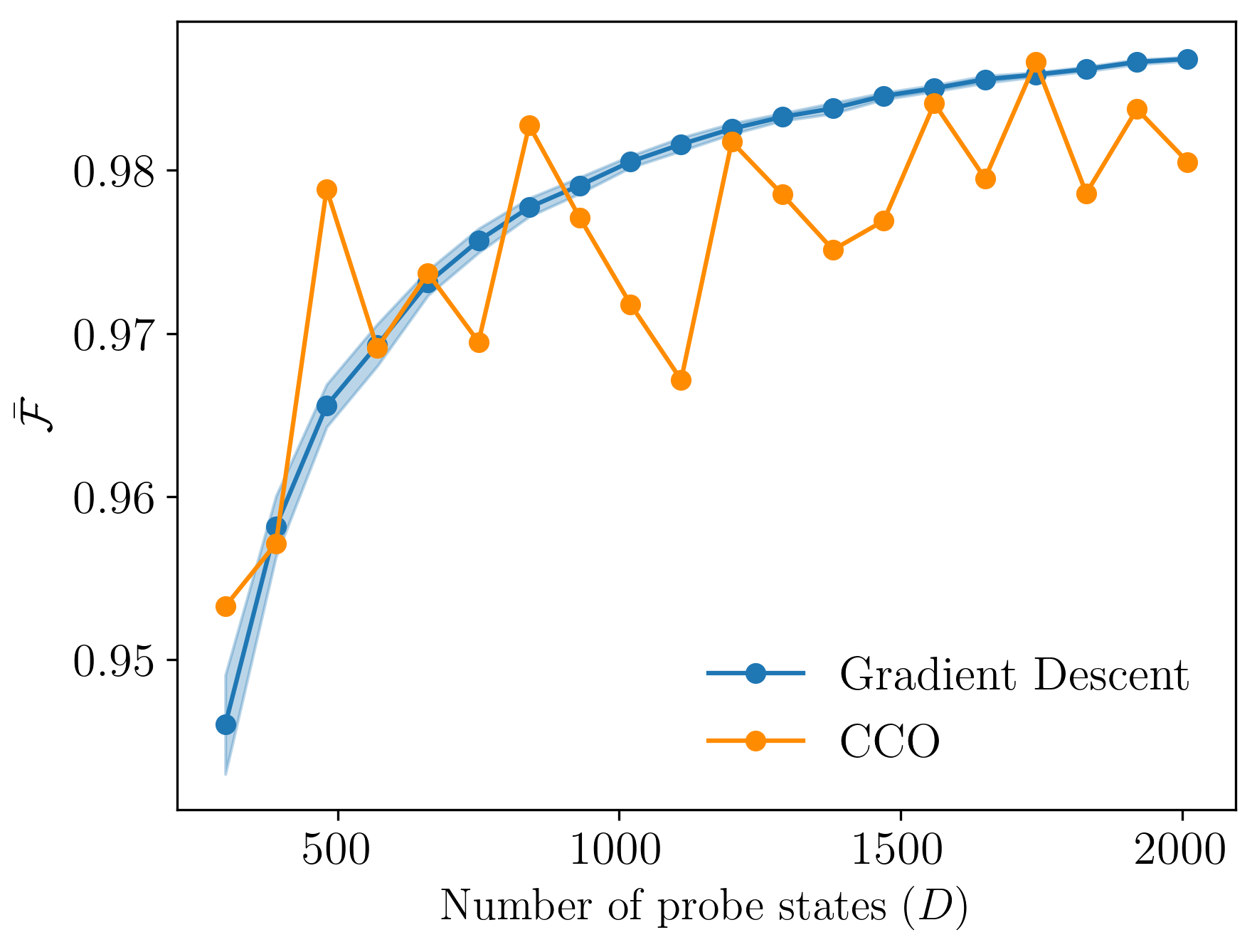}
    \caption{}
     \label{fig:lossy_data}
    \end{subfigure}
    \caption{Average reconstruction fidelity of QDT methods when using datasets of varying size. (a) Average POVM reconstruction fidelity for ideal PNR detector. (b) Average POVM reconstruction fidelity for PNR detector with $85\%$ efficiency.}
    \label{fig:limited_data}
\end{figure}

\begin{figure}
    \centering
    \begin{subfigure}[b]{0.475\textwidth}
    \centering
    \includegraphics[width=1.025\columnwidth]{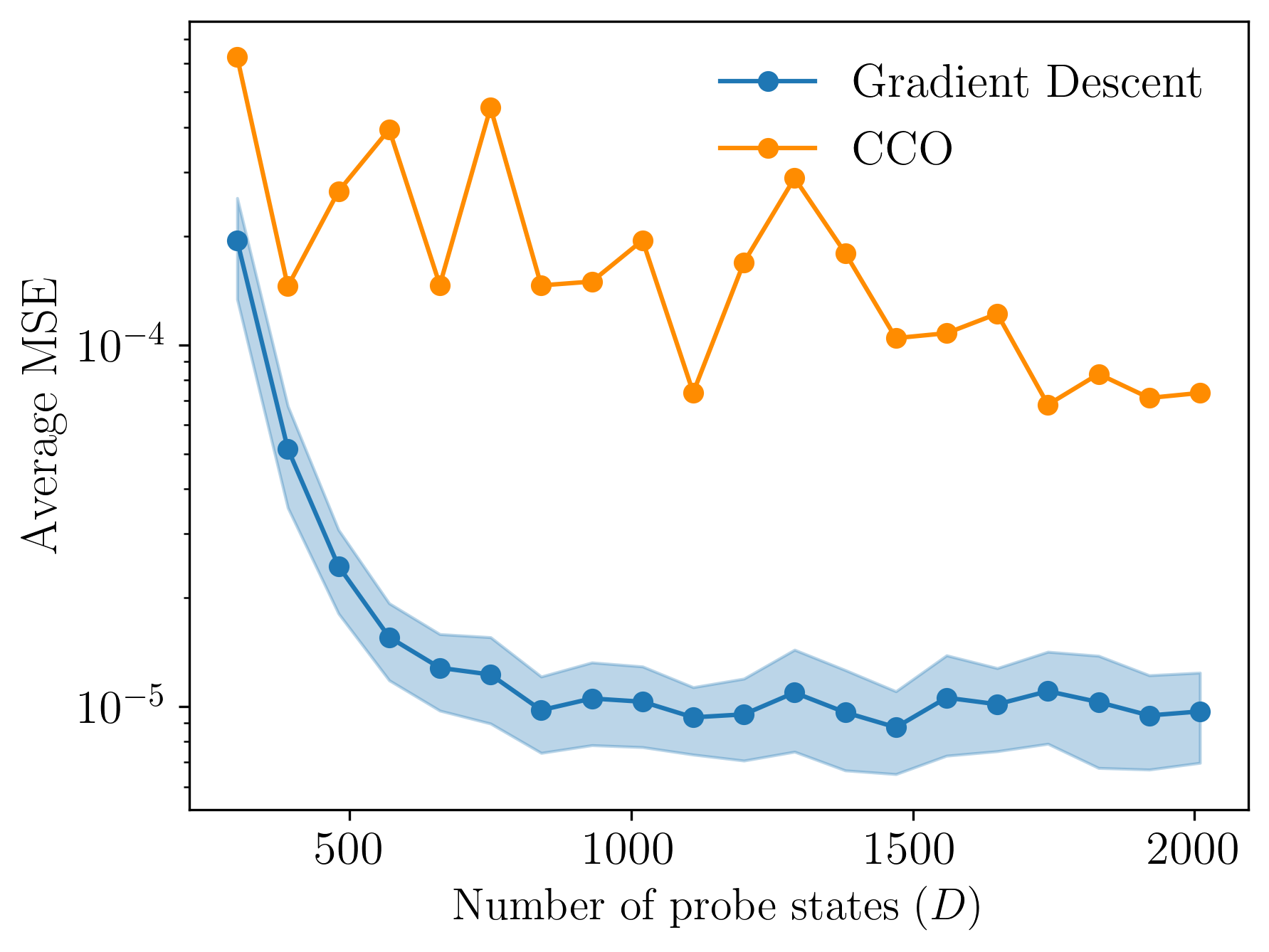}
    \caption{}
    \label{fig:ideal_data}
    \end{subfigure}
    \begin{subfigure}[b]{0.475\textwidth}
    \centering
    \includegraphics[width=1.025\columnwidth]{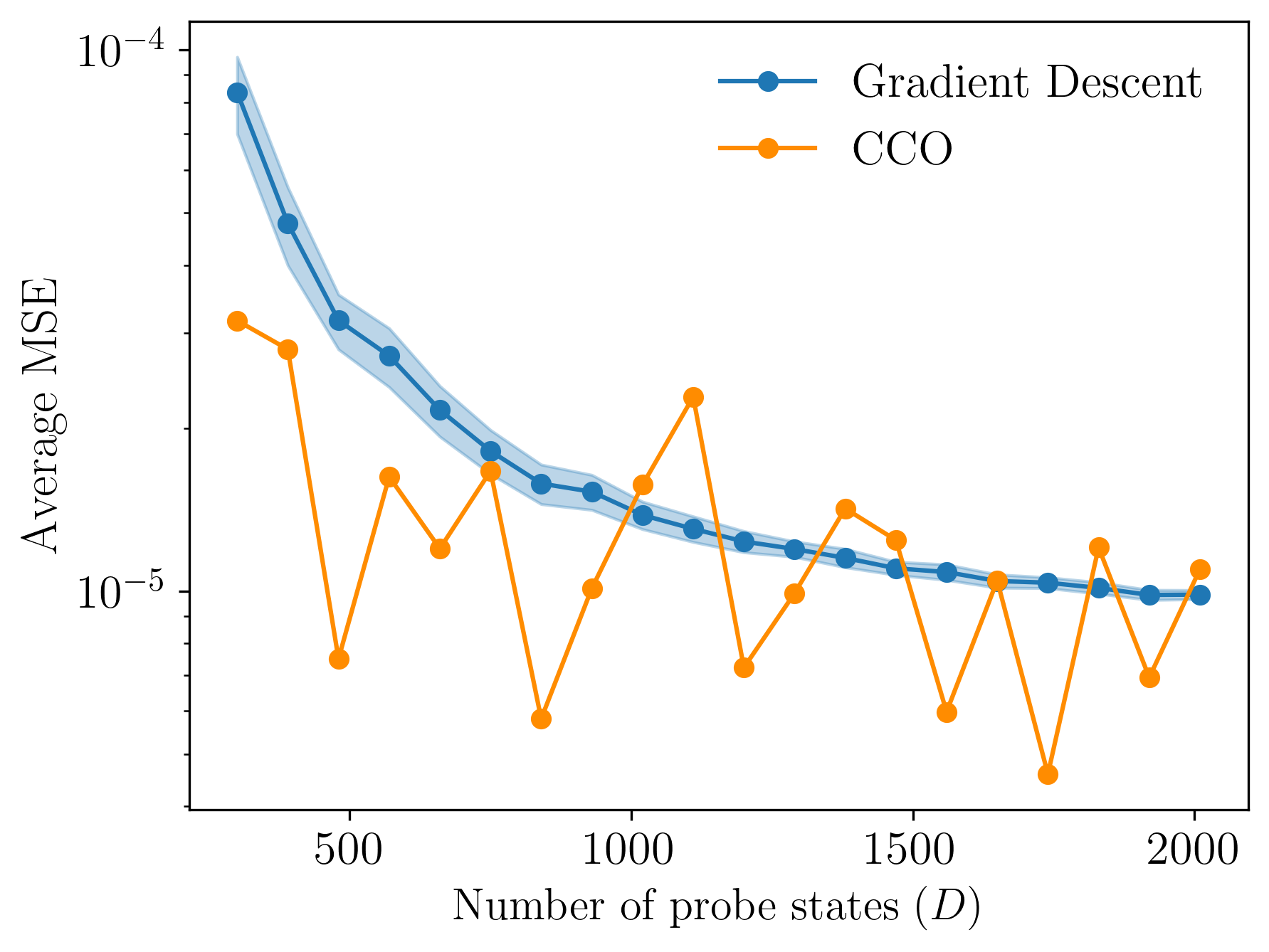}
    \caption{}
     \label{fig:lossy_data_mse}
    \end{subfigure}
    \caption{Average MSE of QDT methods when using datasets of varying size. (a) Average POVM MSE for ideal PNR detector. (b) Average POVM MSE for PNR detector with $85\%$ efficiency.}
    \label{fig:limited_data_mse}
\end{figure}

Here we examine the impact of reducing the dataset size (number of coherent state probes) on the performance of the two methods. This can occur when it is difficult to acquire a tomographically complete dataset (whose size scales like $M^2$ in general) due to the large size of the Hilbert space for high dimensional systems. Fig. \ref{fig:ideal_data} shows the performance of CCO and gradient descent when reconstructing the POVM of an ideal PNR detector with varying numbers of probe states. We can see that gradient descent outperforms CCO for all but the two smallest of dataset sizes. When using the smallest amount of probe states CCO performs slightly better but very similarly in terms of average reconstruction fidelity to gradient descent. 
We can also see that the variance in $\bar{\mathcal{F}}$ decrease with the size of the dataset used for gradient descent.
Fig. \ref{fig:lossy_data} shows the results for an inefficient PNR detector. In this case CCO outperforms gradient descent in more instances but in most cases still achieves lower reconstruction fidelity.
Interestingly, the MSE achieved by gradient descent in Fig.\ref{fig:ideal_data} decreases by about an order of magnitude as more data is added while the CCO method does not achieve such a decrease in error. For the lossy detector case the two methods achieve comparable MSEs as shown in Fig. \ref{fig:lossy_data_mse}.

\subsection{Multi-qubit measurements in the computational basis}{\label{sec:dv}}
For quantum computers implemented using the discrete degrees of freedom (qubits) of superconducting circuits one of the operations most prone to errors is measurement~\cite{tannu2019mitigating}.
In measurement error mitigation protocols for these systems a common simplifying assumption that is made is that the POVM elements describing the detectors are diagonal in the computational basis~\cite{hamilton2020scalable, bravyi2021mitigating, geller2020rigorous, jattana2020general, nation2021scalable, malik2025coherence}.
% so descibes things like classical, i.e. doesn't depend on coherences, cross talk between qubits. It can actually descibe alot of types of both single qubit measurement error and multi qubit (cross talk) measurement error see Ref. `geller2020rigorous`.
This allows for the direct application of our tomographic approach to characterize these detectors. 
In this section we apply our method to reconstruct random $n$-qubit measurements whose POVM is diagonal in the computational basis. 
These POVMs can model various qubit crosstalk errors~\cite{geller2020rigorous}.
For all experiments in this section we use the same hyperparameters from Table \ref{tab:hyperparam_table} and run the gradient descent algorithm 10 times for each POVM and plot the mean and standard deviation of these run.
In all cases we use the tomographically complete set of input states from Eq. \ref{eq:diag_probes_dv}.
\subsubsection{Time complexity}
\begin{figure}
    \centering
    \begin{subfigure}[b]{0.475\textwidth}
    \centering
    \includegraphics[width=1.025\columnwidth]{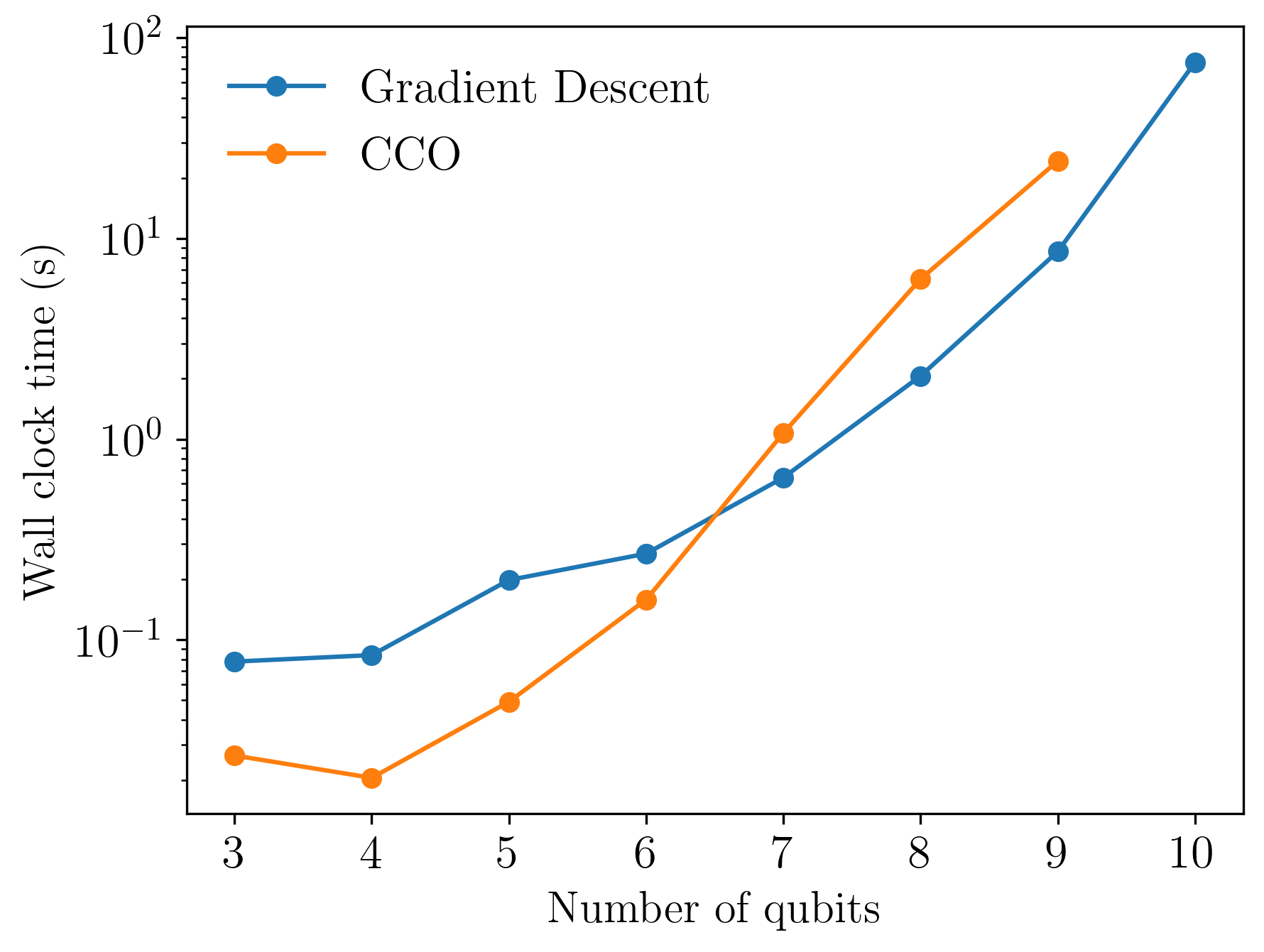}
    \caption{}
    \label{fig:dv_wall_clock}
    \end{subfigure}
    \begin{subfigure}[b]{0.475\textwidth}
    \centering
    \includegraphics[width=1.025\columnwidth]{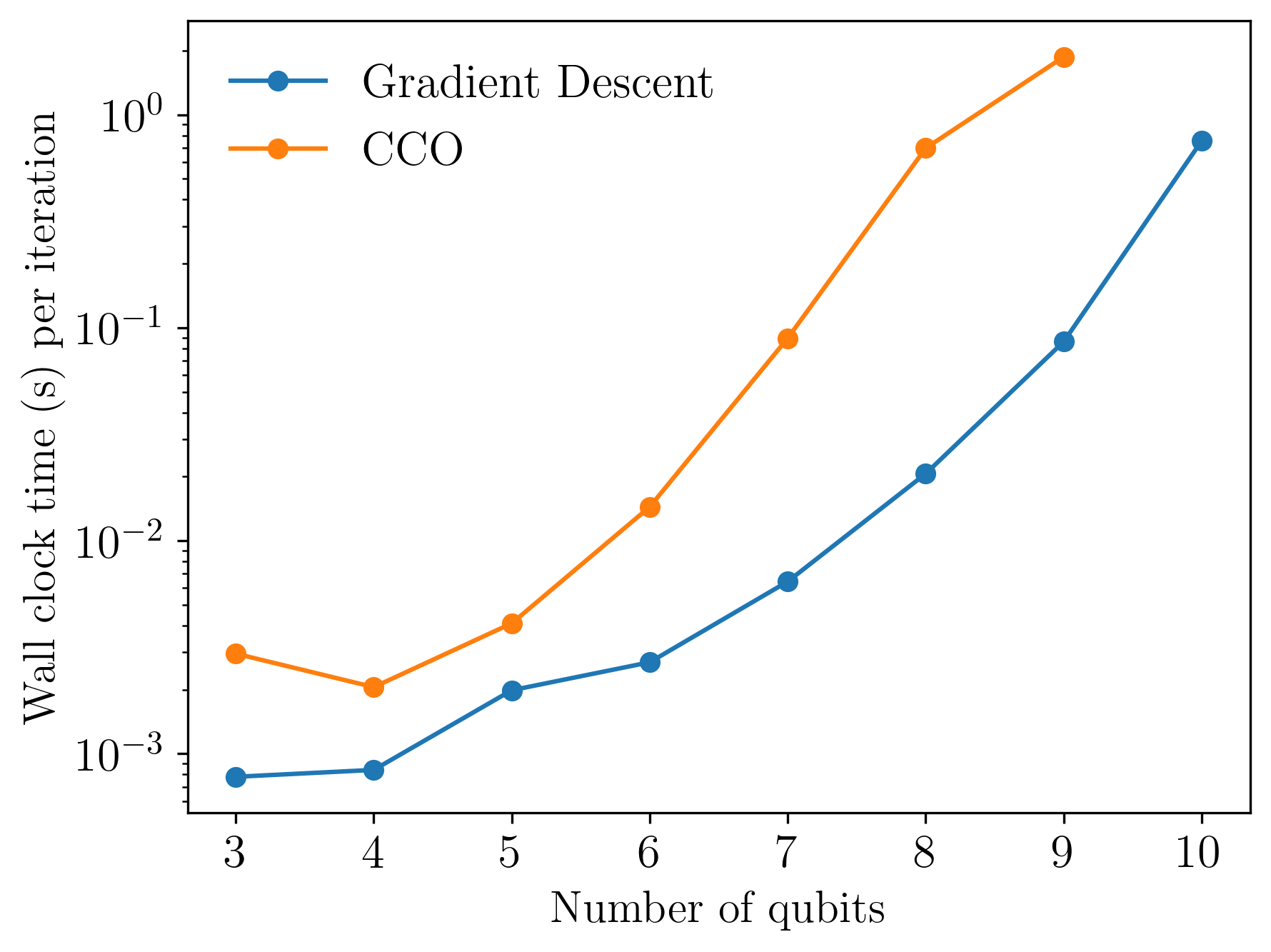}
    \caption{}
     \label{fig:dv_wall_iter}
    \end{subfigure}
    \caption{Wall clock time (in seconds) and wall clock time per iteration for QDT methods for random diagonal $n$-qubit measurement POVMs. Note that for $n=10$ qubits the CCO method fails to find a solution with the memory resources available.}
    \label{fig:time_comp_dv}
\end{figure}

\begin{figure}
    \centering
    \begin{subfigure}[b]{0.475\textwidth}
    \centering
    \includegraphics[width=1.025\columnwidth]{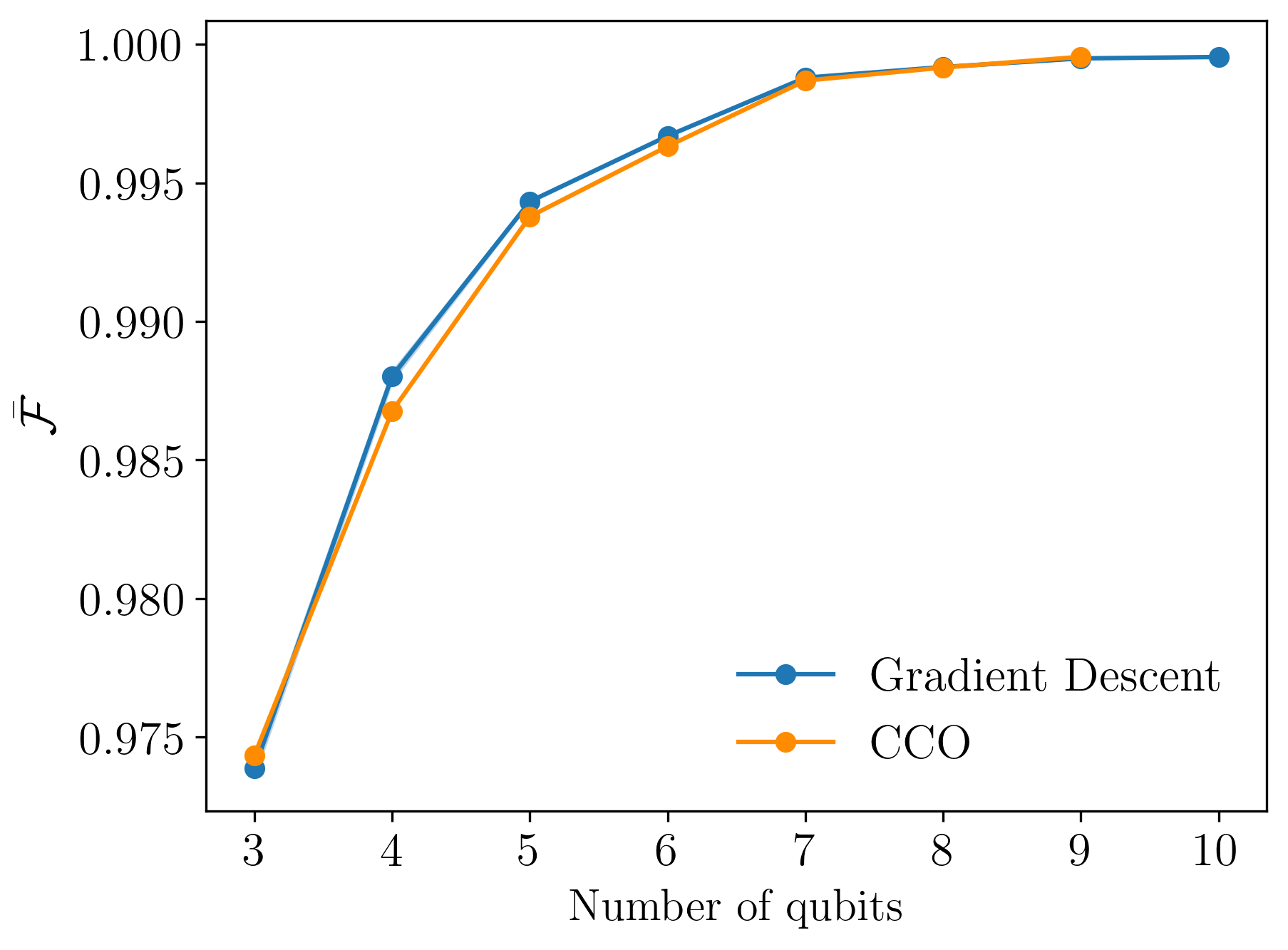}
    \caption{}
    \label{fig:dv_fidelity}
    \end{subfigure}
    \begin{subfigure}[b]{0.475\textwidth}
    \centering
    \includegraphics[width=1.025\columnwidth]{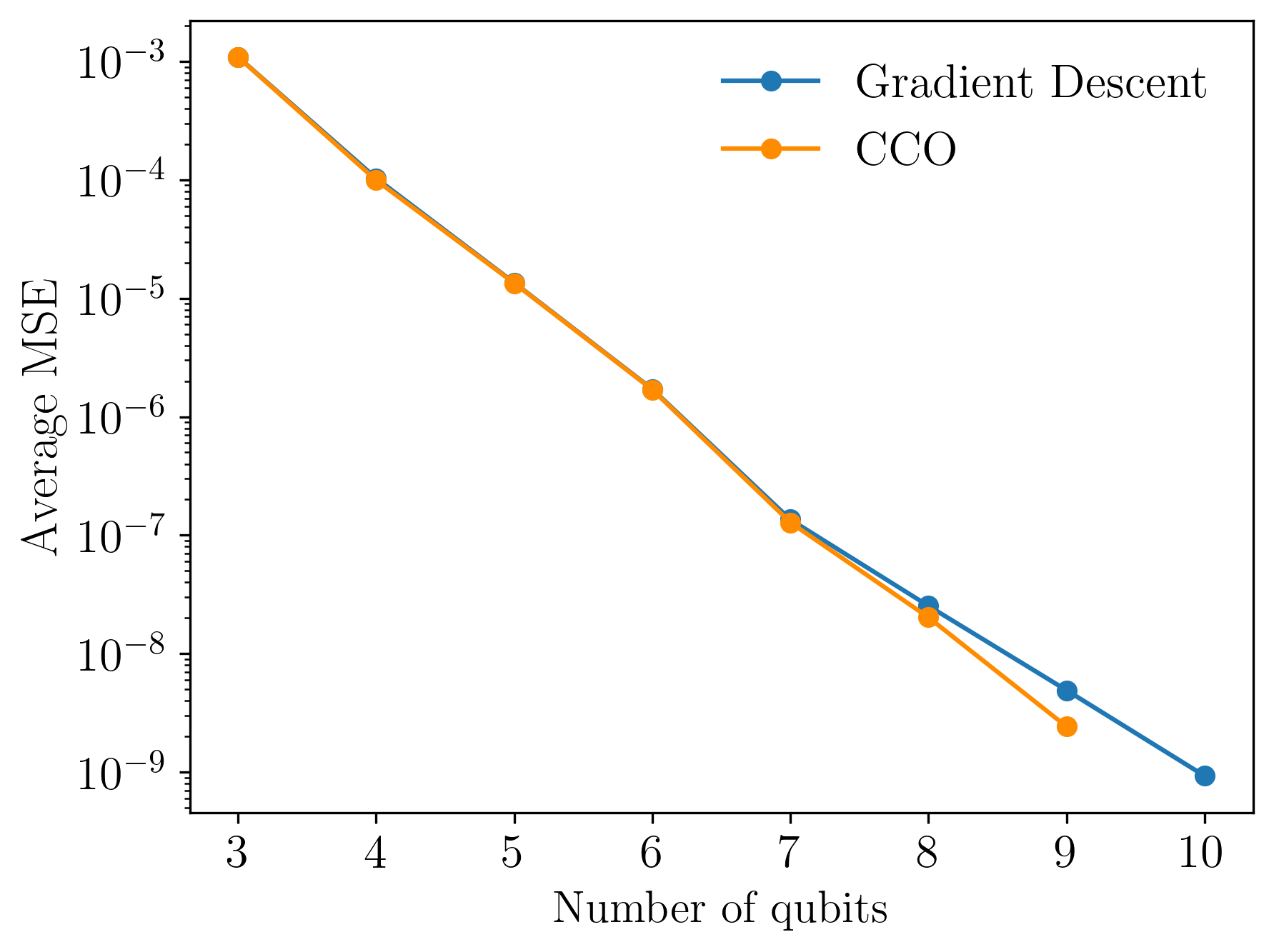}
    \caption{}
     \label{fig:dv_mse}
    \end{subfigure}
    \caption{Average reconstruction fidelity of QDT methods for reconstructing random diagonal $n$-qubit measurement POVMs. (a) Average POVM reconstruction fidelity for the two methods. (b) Average POVM reconstruction MSE for the two methods. Note that for $n=10$ qubits the CCO method fails to find a solution with the memory resources available.}
    \label{fig:fidelity_dv}
\end{figure}

Fig. \ref{fig:time_comp_dv} shows the total wall clock time and wall clock time per iteration of the two QDT algorithms when reconstructing POVMs for various systems sizes.
Note that for systems of more than 9 qubits the CCO methods memory usage exceeds that of our machine and thus does not return a solution.
We can see in Fig. \ref{fig:dv_wall_clock} that for systems with less than 7 qubits the CCO method has lower wall clock time than the gradient descent method. However, the wall clock time per iteration for the CCO algorithm is larger for all system sizes as shown in Fig. \ref{fig:dv_wall_iter}.
Fig. \ref{fig:fidelity_dv} shows the average reconstruction fidelity and MSE of the two QDT methods.
We can see that both methods provide very similar reconstruction accuracies.
\subsubsection{Experimental noise}
\begin{figure}
    \centering
    \begin{subfigure}[b]{0.475\textwidth}
    \centering
    \includegraphics[width=1.025\columnwidth]{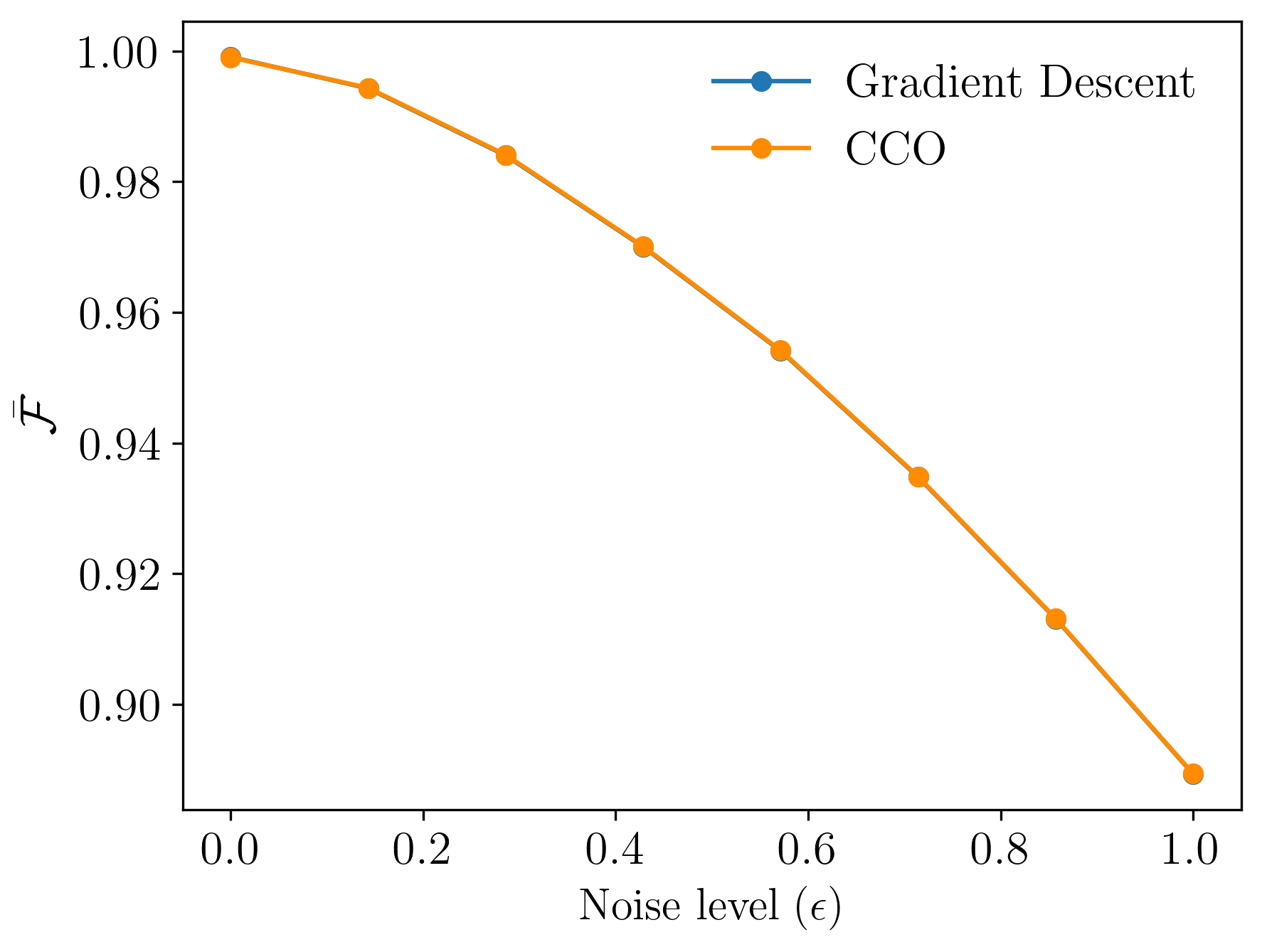}
    \caption{}
    \label{fig:dv_noise_fidelity}
    \end{subfigure}
    \begin{subfigure}[b]{0.475\textwidth}
    \centering
    \includegraphics[width=1.025\columnwidth]{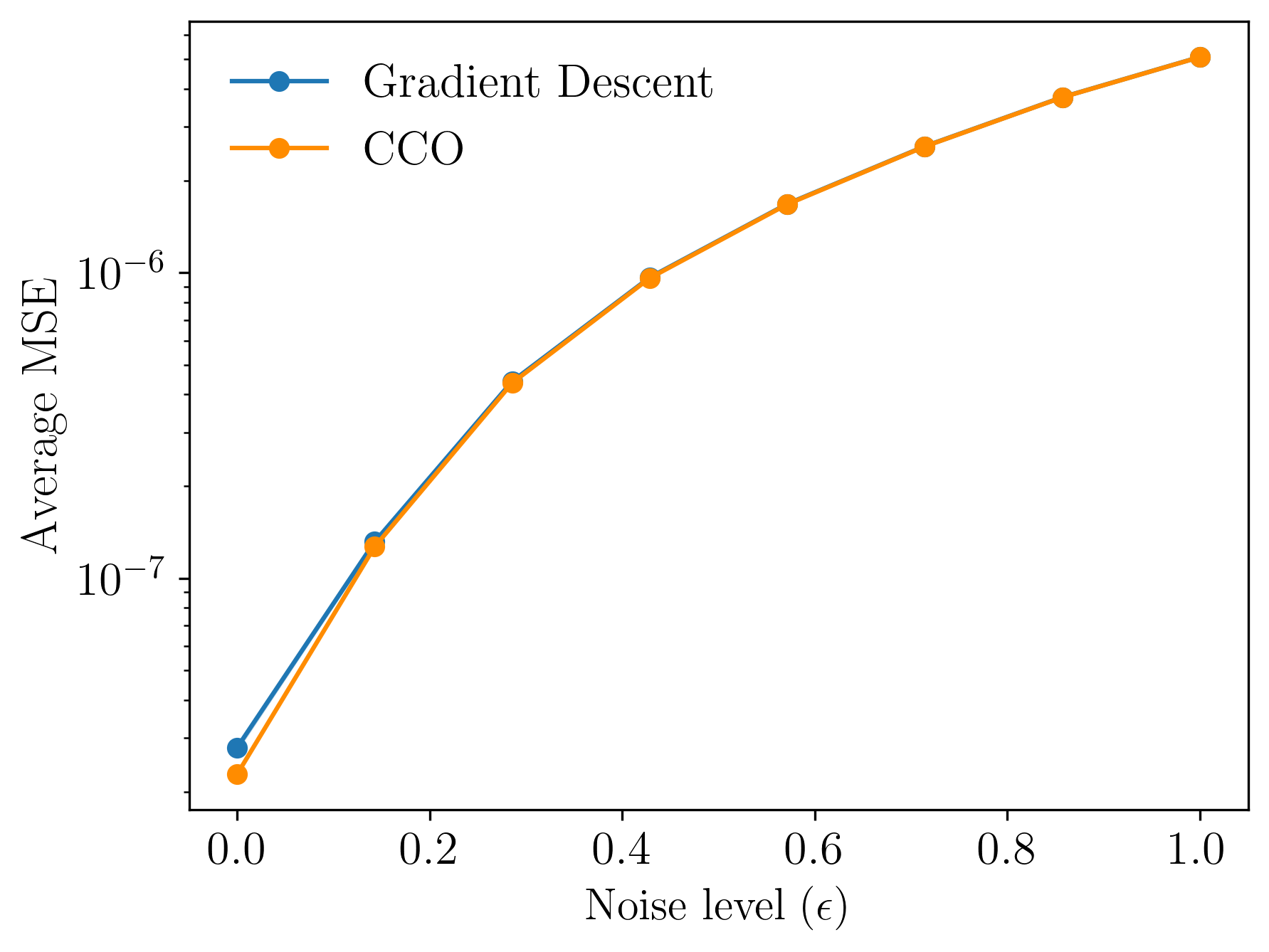}
    \caption{}
     \label{fig:dv_noise_mse}
    \end{subfigure}
    \caption{Average reconstruction fidelity of QDT methods under varying levels of depolarizing  noise in the qubit probe states for random diagonal $8$-qubit measurement POVM. (a) Average POVM reconstruction fidelity for the two methods. (b) Average POVM reconstruction MSE for the two methods.}
    \label{fig:noise_dv}
\end{figure}
In an actual experiment, as in the coherent state probe case, there will likely be some noise in the state preparation process that makes the probes states used to generate the data in $P$ deviate in some way from the ones assumed in Eq. \ref{eq:diag_probes_dv}.
Here we investigate the effect that depolarizing noise in the probe states has on the reconstructions accuracy of the two QDT methods. 
As is well known \cite{guo2023noise} for an $n$-qubit density matrix $\rho$ a depolarizing channel with strength $\epsilon$ transforms the state as 
\begin{align}
    \rho \mapsto (1-\epsilon)\rho + \frac{\epsilon}{2^n}\mathbb{I}_{2^n}.
\end{align}
Fig. \ref{fig:noise_dv} shows the reconstruction fidelity and MSE for reconstructing random 8-qubit diagonal POVMs. 
Interestingly, both methods achieve near identical performance for all noise levels with the performance of both degrading as more noise is present. 
This further shows that gradient descent QDT methods can match the performance of CCO based ones.

\section{Extension to phase sensitive detectors by parameterization of POVMs on the Stiefel manifold}\label{sec:SM_param}
% \textcolor{ForestGreen}{While the method presented above is applicable to many detectors it is useful to extend the generality of the approach to consider phase sensitive detectors.}
Gradient descent based optimization can also be applied to learn POVMs for detectors which are phase sensitive, and thus whose POVMs are no longer diagonal in the Fock basis, via performing the optimization on a Riemannian manifold. %We present a possible approach here and leave the implementation and evaluation to future work. 

%\subsection{Parameterization of POVMs on the Stiefel manifold}\label{sec:SM_param}

The Stiefel manifold has been used for optimization problems in quantum physics as it provides a way to enforce completeness constraints when utilizing gradient descent \cite{ahmed2023gradient, gaikwad2025gradient, luchnikov2021qgopt}. 
In particular Ref. \cite{luchnikov2021qgopt} presents a Python library for performing quantum tomography using Riemannian manifolds however, as noted in Ref. \cite{gaikwad2025gradient}, it does not present a specific parameterization of quantum objects such as density matrices and POVMs on the Stiefel manifold. 
Here we present a novel way of parameterizing POVMs as points on the Stiefel manifold.
% luchnikov2021qgopt does not give specfic paramterizations of POVMs on the Stiefel manifold as stated even in gaikwad2025quantum.
Firstly we will factorize each POVM element $E_n$ into a form that guarantees they remain positive semidefinte and Hermitian by factorizing them, via a Cholesky decomposition, as \cite{benenti2019principles}

\begin{equation}
E_n = W_n^\dag W_n
\end{equation}
where $W_n \in \mathbb{C}^{M\times M}$. We then vertically stack these $W_n$ into a matrix

\begin{align}
\mathcal{W} &= 
\begin{bmatrix}
W_1  \\
\vdots \\
W_N
\end{bmatrix} \in \mathbb{C}^{NM \times M}.
\end{align}
For the POVM described by this matrix to meet the completeness criteria in Eq. \ref{povm_constraints} this matrix must satisfy 

\begin{align}
\mathcal{W}^\dag\mathcal{W} &= \begin{bmatrix}
W_1^\dag, \cdots, W_N^\dag  \end{bmatrix}
\begin{bmatrix}
W_1  \\
\vdots \\
W_N
\end{bmatrix}  
= \sum_{i=1}^N W_i^\dag W_i = \sum_{i=1}^N E_i = \mathbb{I}_M.
\end{align}
Since the matrix $\mathcal{W}$ must satisfy this relation it must be a point on the (complex) Stiefel manifold $St(k, p)$ which for  $k,p \in \mathbb{Z}^+$ is defined as

$$
St(k, p) = \{ X \in \mathbb{C}^{k\times p} | X^\dag X = \mathbb{I}_p \}
$$
which is the set of complex $k \times p$ matrices $X$ that satisfy $X^\dag X = \mathbb{I}_p$ i.e. whose columns are orthonormal. Since we can parameterize our search space as a Riemannian manifold we can also perform optimization constrained onto this manifold using Riemannian gradient descent based optimization methods as done in prior works on QPT, QST and quantum circuit optimization \cite{luchnikov2021qgopt, ahmed2023gradient, gaikwad2025gradient, yao2024riemannian}. 

The Riemannian gradient $\nabla_\mathcal{W}^* \mathcal{L}(\mathcal{W})$ can now be defined, as shown in the supplementary material of \cite{ahmed2023gradient}, via a projection of the Euclidean gradient $ \nabla_\mathcal{W}\mathcal{L}(\mathcal{W})$ onto the tangent space of the Stiefel manifold as

\begin{align}
    \nabla_\mathcal{W}^* \mathcal{L}(\mathcal{W}) = A\left(\mathbb{I} + \frac{\gamma}{2}B^\dag A\right)^{-1} B^\dag \mathcal{W}
\end{align}
where 
\begin{align}
G =\frac{\nabla_\mathcal{W}\mathcal{L}(\mathcal{W})}{||\nabla_\mathcal{W}\mathcal{L}(\mathcal{W})||_2}, \quad A=[G, \mathcal{W}], \quad B = [\mathcal{W}, -G]
\end{align}
and $\gamma$ is the learning rate.
The Riemannian gradient can then be used in the standard way to update the parameters in the steepest direction in parameter space that minimizes the objective function, while remaining on the Stiefel manifold, as

\begin{align}
    \mathcal{W}' = \mathcal{W} - \gamma\nabla_\mathcal{W}^*\mathcal{L}(\mathcal{W}).
\end{align}
An advantage of this formulation is the ability to control the rank of the resulting POVM elements by making the individual $W_i$'s have shape $r_i \times M$ as this results in $E_i$ having maximum rank $r_i$ when $r_i < M$. This allows one to use a rank-controlled ansatz for the tomography and reduces the dimensionality of the problem if prior information about the POVM is known e.g. an ideal PNR detector has POVM elements with rank $r=1$.

\section{Conclusion}
We have proposed a gradient descent based method for performing tomography of arbitrary phase insensitive quantum detectors. This method is more scalable than approaches based on CCO, as benchmarks and complexity analysis show, while also achieving comparable reconstruction accuracy. 
This work also opens up the possibility of augmenting quantum detector tomography with the heavily researched toolkit developed for neural network training. 
This includes scaling to large systems (Hilbert space dimension and detection outcomes) or large datasets (number of probes states) using the \texttt{PyTorch} distributed data parallel module to leverage multiple GPUs \cite{li2020pytorch}, as well as improving memory and time efficiency during optimization through methods such as automatic mixed-precision training \cite{narang2017mixed}. 
Additionally, as done in Ref. \cite{schapeler2024scalable}, one can exploit the sparse and banded nature of the matrix $F$ that arises when $M$ is large to greatly reduce both memory and time requirements by using the sparse storage and efficient sparse-dense matrix multiplication routines present in \texttt{PyTorch}.

We have also presented a way to generalize of our scheme to arbitrary detectors which are not phase insensitive. 
The implementation and evaluation of this more general scheme is left for future work.
A question deserving further investigation is if one can incorporate the normalization condition of the $\Pi$ matrix rows in a way that preserves convexity of the original objective function. 
This would be useful as gradient descent has much stronger convergence guarantees when the objective function is convex \cite{bubeck2015convex}. Lastly the application of gradient descent to solve the optimization problem used to perform QDT for arbitrary phase sensitive detectors in Refs. \cite{zhang2012mapping, zhang2012recursive} would be another research direction of interest that could bypass the need for Riemannian manifold based optimization. 

\textit{Note}: After our manuscript appeared (arXiv:2511.14579), a work was posted (arXiv:2511.15682) which also introduced methods for performing QDT using gradient descent with the same Stiefel manifold parametrization presented in Section \ref{sec:SM_param} as well as a different approach based on a Hermitian operator normalization via eigenvalue-scaling technique~\cite{gaikwad2025quantum}. 
We compare and contrast the two approaches in the Appendix.

\begin{acknowledgements}
OP was supported by U.S. National Science foundation grants OSI-2531569 [NQVL-QCAP NQVL:QSTD:Design: Quantum Computing Applications of Photonics (QCAP)], PHY-2514971, and ECCS-2530171.
\end{acknowledgements}

\appendix

\section{Comparison to other gradient-based detector tomography methods}{\label{appendix:compare}}
After our manuscript appeared (arXiv:2511.14579), a work was posted (arXiv:2511.15682) which also introduced methods for performing QDT using gradient descent with the same Stiefel manifold parametrization presented in Section \ref{sec:SM_param} as well as a different approach based on a Hermitian operator normalization via eigenvalue-scaling technique% (HONEST)
~\cite{gaikwad2025quantum}.
While these approaches can model phase sensitive detectors, they are less efficient for the diagonal POVM case as they reconstruct the dense POVM matrices by default leading to space and time complexity of $O(NM^2)$, in addition to needing to compute matrix inverses for the Stiefel manifold method and eigenvalue decompositions for the eigenvalue-scaling technique, %HONEST method, 
as opposed to $O(NM)$ for our method which requires neither the computation of matrix inverses or eigenvalue decompositions.
Note that three out of the five POVMs studied in the work (PNR detectors, bucket detectors and Pauli-Z projective measurements) have POVMs which are diagonal and thus can more efficiently be tomographed using our method.
The work also presents the same rank controlled ansatz as in Section \ref{sec:SM_param}. 
While this ansatz can, in principle, be used to reduce the number of parameters to be optimized down to $O(NM)$, this is only the case when all POVM elements have rank 1, which is not always the case for diagonal POVMs. 
Two examples of this are \textit{(i)}, the POVM elements for a PNR detector, which are diagonal but still full rank when detector efficiency is not unity and \textit{(ii)}, the POVM element of a bucket detector that corresponds to light detection which has rank $M-1$ even in the regime of perfect detector efficiency. 
We also note that, even in the case of using a rank-1 ansatz, reconstructing a diagonal POVM with the methods presented in \cite{gaikwad2025quantum} requires constructing the dense POVMs to evaluate the loss function as can be seen in the publicly available code provided by the authors. 
% See line 396 in https://github.com/agtomo/SGD-QMT/blob/main/code%20and%20tutorial/SGD_QMT.py#L396
The method presented in Ref.~\cite{luchnikov2021qgopt} also suffers from the same problem when working with diagonal POVMs and the time complexity of $O(NM^2)$ for POVM reconstruction is explicitly stated in Table 3 of Appendix B (note that they use $N$ to denote the Hilbert space dimension and $M$ to denote the number of POVM elements). 
Our method, in contrast, retains $O(NM)$ complexity in all of these cases.

\bibliography{Anteneh,Pfister}% Produces the bibliography via BibTeX.

\end{document}